\documentclass[12pt, titlepage, twocolumn]{report}

\usepackage{titlesec}
\usepackage{fancyhdr}
\usepackage{lipsum}
\usepackage{tikz}
\usepackage{pgfplots}
\usepackage{siunitx}
\usepackage{pgfplotstable}
\usepackage{graphicx}
\usepackage{mathtools}
\usepackage[printwatermark]{xwatermark}
\usepackage{xcolor}
\usepackage[labelfont=bf]{caption}
\usepackage[T1]{fontenc}
\usepackage{blindtext}
\usepackage{amssymb}
\usepackage{amsmath}
\usepackage{algpseudocode,algorithm,algorithmicx}
\usepackage{fancyvrb}
\usepackage{geometry}
\usepackage[utf8]{inputenc}
\newtheorem{theorem}{Theorem}
\usepackage[draft]{minted}
\usepackage{listings}
\definecolor{lightgray}{rgb}{.9,.9,.9}
\definecolor{darkgray}{rgb}{.4,.4,.4}
\definecolor{idk1}{rgb}{.4,.6,.8}
\definecolor{purple}{rgb}{0.65, 0.12, 0.82}

\lstdefinelanguage{js}{
  keywords={import, Blockchain, Consensus, Root, Mechanism},
  keywordstyle=\color{blue}\bfseries,
  keywords=[2]{AddRoot, func},
  keywordstyle=[2]\color{idk1}\bfseries,
  keywords=[3]{Aspect, AddAspect},
  keywordstyle=[3]\color{red}\bfseries,
  identifierstyle=\color{black},
  sensitive=false,
  comment=[l]{//},
  morecomment=[s]{/*}{*/},
  commentstyle=\color{purple}\ttfamily,
  stringstyle=\color{red}\ttfamily
}

\graphicspath{ {./images/} }

\titleformat{\section}
{\normalfont\scshape\centering}{\thesection}{1em}{}

\titleformat{\section}
{\normalfont\Large\bfseries}{\thesection}{0em}{}[{\titlerule[0.8pt]}]

\titlespacing\section{0pt}{12pt plus 4pt minus 2pt}{10pt plus 2pt minus 2pt}

\date{\today}

\date{}

\titleformat{\section}
  {\normalfont\scshape\centering}{\thesection}{1em}{}
  
\titleformat{\section}
  {\normalfont\Large\bfseries}{\thesection}{1em}{}[{\titlerule[0.8pt]}]

\makeatletter
\renewcommand\thesection{}
\renewcommand\thesubsection{\@arabic\c@section.\@arabic\c@subsection}
\makeatother

\begin{document}
\begin{titlepage}

\protect\parbox{.9\textwidth}
	{\protect\centering 
		\huge Exposing A Customizable, Decentralized Cryptoeconomy as a Data Type 
	}
\begin{center}
{\LARGE
\today}
\end{center}

\vfill
{
\Large
\begin{center}
\texttt{The Kunta Project} \\
\texttt{Jovonni L. Pharr} \\
\texttt{[v0.0.1]}
\end{center}
}

\vfill
\section*{Abstract}
\noindent

Purposely modular, this protocol enables customization of several protocol properties, including the consensus properties implemented, blockchain type, the roots used, and virtual machine opcodes, among others. These modules enable implementing parties to control the behavior of their economy, with a minimal amount of effort, and no sacrifice in participant cryptoeconomic quality. This work also demonstrates the simplification of the developer experience by abstracting away all technological details, except basic CRUD-based operations, using various programming languages. We demonstrate the mechanism design approach taken, and formalize a process for deploying populations of blockchain economies at scale. The framework shown includes adequate tooling for simulation, development, deployment, maintenance, and analytic-based decision making. Lastly, we introduce an expressive programming language for the purpose of creating, and interacting with the cryptoeconomy designed by the implementing developer.

\end{titlepage}

\tableofcontents
\listoffigures

\renewcommand\listoflistingscaption{List of Code Snippets}
\listoflistings

\clearpage

\onecolumn
\section{Manifesto}

Though not exclusively the case, the state of the present for this blockchain status quo is a direct reflection of the technological, economical, and game theoretic frameworks of yesteryear. The blockchain ecosystem of today steadily moves towards more generalized, general-purpose architectures aimed at being applicable to as many problem spaces as possible. Typically, in order to alter a blockchain protocol's design, a developer must embark on a reverse engineering effort of which may seem daunting to even seasoned developers. This is a severe hinderance to a subset of the developer community interested in experimenting with building decentralized solutions.

At the time of writing, the standard blockchain protocols and platforms consist of a single designed protocol, upon which one can either deploy a blockchain application, or a private, and dedicated blockchain network -- these protocols address the usability of implementations using blockchain. Developers can create "smart contracts", but this level of development does not include customization of the behaviors of the protocol itself. Developing "smart contracts" is in the service of including additional computations within a blockchain -- not holistically customizing and dictating which computations happen within the chain fundamentally. 

Today, there lacks simplified developer experiences, of which lead to third party wrappers around the protocol to fill these gaps. Cloud computing companies are beginning to offer abstraction layers around protocols as an attempt to offer "enterprise" blockchain solutions. These endeavors still do not offer customization opportunities regarding the protocol. Only A small number of applications where you truly need borderless global censorship resistant neutral platform to establish trust and immutability that cannot be attacked, or compromised by anyone, is not many. However, for the developers who seek it, developer experience should be taken with the highest priority.

By relieving the brain of all unnecessary work, a good abstraction sets us free to concentrate on more advanced problems, and in effect increases the mental power of the participants. It is the democratization of decentralized technologies that is the long-term goal of this project -- not the simple data structure, and concept blockchains are built upon.

\begin{quotation}
There is no room for \(\texttt{Tribalism}\) when the mission is democratization
\end{quotation}

\begin{figure}[ht]
	\begin{flushright}
		\includegraphics[width=0.10\textwidth]{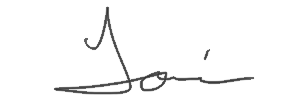}	
	\end{flushright}
\end{figure}

\clearpage
\onecolumn

\part{Preliminaries}
\chapter{State of Today}

\section{Introduction}
The introduction of Blockchain to the world has been covered by writers from several backgrounds; this paper does not aim to serve as yet another proponent of the usage of Blockchain ecosystems. Several implementations of Blockchain systems exist today, and the variation between them can be nonexistent, subtle, or dramatic. One property of the these cryptoeconomic systems is that the development teams tend to believe their implementations are the "best" blockchain implementation for their proposed uses -- of which can be specific, or general. 

These platforms can be purposed for specific solution spaces, such as Bitcoin, and others are more general such as Ethereum. This project is purposed to introduce not only a protocol development platform, but a change in the blockchain development paradigm. Such a change can result in current platforms becoming old-fashionaed, but it is in the spirit of evolution for cryptoeconomic design.

At the time of this writing, there is an unspoken/well-spoken tension existing within the "Cryptocurrency" world. However, these arguments stem from disagreements about protocol design, protocol governance, game theoretic and cryptoeconomic design, or interpersonal discrepancies. These interactions not only display less-than-professional behavior, but also distract the public towards caring more about the implementation details of any blockchain protocol -- of which are arbitrary, and do not universally apply. Being a concept, and design approach, individual blockchain-based implementations should never be communicated as generalized, or universally "correct", as this is only as strong as the assumptions made in designing the implemented protocol itself.

\section{Public Perception}
There are various ways in which the concepts of blockchains are taught, and some closely reflect the truth, others do not; we believe this is due to purposeful simplification of blockchain protocols. It is understood that explaining complexed ideas at a high level works, but there are ramifications with doing so. The individuals being educated, if they are shielded from important details, may miss the "magic", or actual "power" behind the concept.  

This has led to some people denouncing blockchains as nothing but "inefficient databases", "hashed linked lists", and so on. It isn't the case that these are wrong, but these simplifications, abstract away any cryptoeconomic property of using them. To discuss Cryptoeconomic systems without seriously considering how incentives, and economics influence the behavior within it, is a disservice to the discipline.

\subsection{Public confusion}
At the current time of writing, popular culture tends to use the term "The Blockchain" when referring to this space, and its techniques/technologies. However, this convention has led to wide spread confusion, and ignorance. To understand the power behind designing, and deploying a blockchain powered economy, one must at least be introduced to the discipline of Cryptoeconomics, and mechanism design. Referring to a concept, using "the" as a prefacing term, indicates a sole instance of that concept. "The blockchain" implies, there is one, of which people default to being Bitcoin, or Ethereum. It is seldom taught in mainstream as a concept existing in the intersection of cryptography, computer science, and economics.

\section{Current Implementations}
Current implementations of Blockchain networks, from bitcoin, to all platforms at the current time of writing, all have one thing in common: the blockchain structure, mechanisms, virtual machines, and operation codes are all ideated, designed, and developed by the creators of the system. Though the systems are open source, one must reverse engineer the blockchain platform itself, and build from it's source code. The more complexed the open source design is, the higher the knowledge and talent requirement is for anyone wishing to alter its design. At the time of writing, it is uncommon that a decentralized protocol is designed to be customizable, and still maintain a flexible, reliable, and expressive core.

No matter how one perceives the purpose of a Blockchain, one common purpose of any implementation is storage. This is because at any point in time, from the genesis block of any blockchain, there exist \(> 1 \texttt{bit}\) of information within it. Since the platform's developers decide how information is stored, and what to store, we must abide by their design. Platforms do allow developers to store additional information on the blockchain used, but developers have no say so regarding what is fundamentally stored on the blockchain, or how it is stored. 

The argument can be made that the developers have done the R\&D work to decide what "best practices" are, but this results in external developers being "locked out" of the development process; in other words, development tends to be centralized in practice. From a customer experience perspective, this is hard to justify in a world of many talented developers. For platforms that take developer experience into account in the highest regard, this is unacceptable since the platform is to "serve" developers.

Since Blockchains store information fundamentally, we can draw an analogy to database architectures of years past, and present. There exists a plethora of proprietary, and open-source database systems, and paradigms. The evolution of NoSQL \& SQL have allowed developers to choose the schema of their databases, and allow for a wider usage space.

\chapter{Predecessors}

\section{The influence of Bitcoin}
Bitcoin has seemingly convinced cryptoeconomic-enthusiasts that a singular blockchain protocol is the answer to many problems, and this has resulted in a "protocol race" for the supreme, most usable, and most general platform. This is the main motivation for platforms such as Ethereum, Cardano, Wave, and a host of others. This assumption has led developers to compete regarding their blockchain ecosystem designs. Bitcoin can be viewed as a transaction-based state machine, whereby transactions in the network progress the protocol's state from \(\sigma_t \rightarrow \sigma_{t+1}\) through \(\Upsilon\), the state transition function.

\begin{equation}
\sigma_{t+1} \equiv \Upsilon (\sigma_t,T)
\end{equation}

A state transition function, \(\Upsilon\), can be thought of as Alice sending \$1 to Bob, and that "sending" action transitions the system's state from Alice having \(\$x\), to Alice having \(\$x-1\), and Bob having \(\$y + 1\); Bitcoin can be viewed this same way using this analogy.

\subsection{Mechanism in Bitcoin}
Bitcoin, by adopting more interest has raised concern regarding fairness across the network. Users cannot simply mine on their Personal Computers these days,

Bitcoin's influence has resulted in large organizations, and groups, along with dedicated hardware, focusing resources on securing the network. The difference in resources between these groups, and the average interested party, can be seen as unfair. Some of these groups are even trashing their units. In addition, some speculators believer a small group of entities own a large subset of Bitcoin. This can even been seen by exploring for the "richest" accounts on the network. If true, hinders promotion of the project, and threatens the very democratic ethos of Bitcoin's vision. 

The Coase theorem states, with zero or little transaction cost and unambiguous rights to assets, the market could reach what is know as "Pareto efficiency" regarding the allocation of resources, despite to whom the property is allocated. 

The Bitcoin project initially attempted two approaches to creating a "fair" system: 

\begin{enumerate}
	\item allocating Bitcoins to all users according to the number of the nodes, namely one-IP-address-one-vote; 
	\item allocating Bitcoins to the miners according to the computing power, namely one-CPU-one-vote.
\end{enumerate}

 Due to security issues concerning users being able to easily access many IP Addresses, one-CPU-one-vote began the approach after release. The Miners processing transactions provides a fundamental incentive mechanism for the network because Miners receive "reward" for doing so. The total computing energy created by miners served as a fundamental security measure, supporting activities (transactions) by non-miners.

Miners also pay computing resources to users, as a total, of the network, in exchange for ownership of Bitcoin. This is viewed as transaction fees, but support the network as another incentive mechanism. If the computing power expened by miners ends up costing more than the revenue generated from mining, miners would be incentivized to offer as much computing power. The bigger the difference between mining costs, and user-purchasing costs, the higher the incentive for additional miners to join the network becomes.

\subsection{Altcoins}

\begin{figure*}[ht]
\centering
	\includegraphics[width=\textwidth]{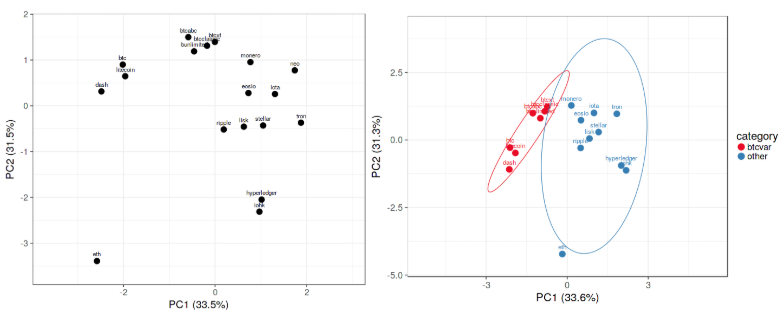}
	\caption{6 dimensional GIT behavior visualized using principal component analysis. Right: (Red) Bitcoin Variant GIT Projects. (Blue) Non-Bitcoin Variant projects.}
	\label{git_ana_both}
\end{figure*}

Alternative coins have been introduced since Bitcoin, the majority of which are direct GIT clones of the original Bitcoin source code; this has further pushed the ethos of "Best Protocol" wins by project teams developing on top of it.

This cultural ethos again embodies evolution towards either designing a protocol for a specific use case, of which may not work applied to anything else, or the most robust, multipurpose protocol design.

In figure \ref{git_ana_both}, we use 6 GIT metrics to analyze blockchain projects. We begin with 500 tokens from the Coin Market Cap (CMC) index, and focus on the top 17 projects for clarity. The GIT metrics used:

\begin{enumerate}
	\item Commits
	\item Contributors
	\item Number of Repositories
	\item Number of Github Organization Users
	\item Number of Releases
	\item Number of Branches
\end{enumerate}

For metrics regarding a single repository, we simply use the most active, and most used/active repository within the project's GIT organization. For example, for the Ethereum organization, we use their C++ Aleth project, of which has the highest amount of commits under their organization, and is still active. This approach on the metrics to use from GIT is arbitrary, and not quantified beyond this. We perform the same logic for each of the 500 tokens. 

In figure \ref{git_ana_both}, we use Principal Component Analysis (PCA) to view GIT activity. In the right plot in figure \ref{git_ana_both}, the red cluster contains projects that are forked from Bitcoin's source code (called "Bitcoin Variants"), and the blue cluster contains projects not forked from Bitcoin. Clearly, a cluster emerges for projects forked from Bitcoin, based on their GIT metrics (which isn't surprising given the fact that GIT metrics transfer with the forking of a project), we claim GIT metrics provide meaningful insight for determining the behavior of a project, project team, and inheritance of the repository. This is also the case when using the 500 total tokens on CMC.

\subsection{Protocol Debates}
At the time of writing, amongst the thought leaders in the decentralized technology space, there exists a theme of argumentation over protocol design. 

\subsubsection{Block Size}
A famous, and ongoing debate exists regarding the preferred block size in the Bitcoin network. There have been many hard forks of the original Bitcoin source code, and has resulted in a myriad of schools of thought. Roger Ver, a well-known thought leader, and early investor in Bitcoin-based companies argues, sided with Satoshi Nakamoto, that increasing the block size to whatever it is required to be, is a feasible long term solution due to "Moore's Law". However, research strongly indicates the trend of Moore's Law slowing down, and may come to a halt. This is due to the space constraints with creating more transistors fitting in a predefined space. It not only becomes more difficult to practically place more transistors in a space, but their exists other physics based problems with creating ever smaller logic gates. See CITE. 

Simply increasing the block size is a naive approach to better design, and wider adoption. Moreover, the requirement to choose ever more clever methods, some of which change a fundamental behavior of the system, all to see the success of a protocol, is more of a design shortcoming on the shoulders of the protocol engineers, and less of an economic property. This is a direct ramification of develop a one-size fits all protocol -- though it was specific to the finance industry. 

\subsubsection{Source Code Ownership}
Since the original Bitcoin source code repository has changed ownership, controversy has emerged regarding what the correct design choices should be moving forward. This has brought about many proposal towards improving Bitcoin. These proposals are decided upon not by the author of the original Bitcoin code, but the owners of the repository at the time -- not to mention it being "forkable", since it is open sourced. A notable "improvement" was moving from "First Seen First Safe", the transactions that are seen first, get included into the block, to "replace by fee", where you can replace a seen transaction with a different transaction up until any point it is included in a block. This went from transactions being basically "instant", due to being "unremovable" as soon as it's seen by the protocol, to being able to be replaced based on the amount of the transaction fee relative to other transactions.

This is a direct ramification of a lack of direct guidance under an open source project; there are ramifications to open-sourced software. However, the public, or "participating" public, should always have the tools available to verify every aspect of any blockchain implementation, or else the protocol is useless in the decentralized space.

\subsection{Global Attack Resistance}
An aspect of Bitcoin's story, that often goes unnoticed, is the effect of the project "flying under the radar" for a number of years. By the time Bitcoin reached mainstream attention, the network already built a history of hashing power performed by the miners securing the network. For any blockchain based cryptoeconomy using a security mechanism similar to Proof-of-Work, to scale today, the implementation must reach "scale" before it is attacked "at scale". In other words, it must possess a strong enough economic base to resist attack, also known as the "security margin" of a system. If the network is attacked at a lesser scale than the scale of its security margin, is may withstand attack.

Any motivated attacker with enough hashing power, can create an alternate history of the chain, built from the same genesis block, that has a longer, and still "valid" chain; This can be difficult to deal with for any protocol implemented. If you have a innovative consensus algorithm, and people think its going to be valuable, and think it will be valuable enough to "further", they may also believe it is valuable enough to attack. Any newly implemented protocol, or network, may not have a blockchain with a high enough security margin to resist a coordinated attack. The notion that Bitcoin has withstood attacks is an important property of its implementation, and enables many of its benefits.

\subsection{Process Consensus}
Process consensus is a process of debate and proposal that occurs in the development community. On several platforms (mainly a code repository), users submit improvement proposals, or "modifications" to the rules of the system. By gathering, debating, and trying to reach a process consensus, if enough people agree with the proposal, it may be taken further for performance testing. Once a user, or users provide a reference implementation that demonstrates the change, some performance analysis on a "Test Network", and an iterative developer review, it is eventually implemented in the core reference implementation -- this is called reference consensus. In order for the software to propagate, nodes must upgrade, and this requires consensus among the constituencies.

\subsection{Ossification}
After a while, the protocol gets embedded into many implementations, and thusly becomes harder to upgrade/change. This is similar to the effects of the ossification of IPV4, which caused the long term effort to upgrade to IPV6. Since protocol innovation becomes harder over time, proposed solutions require innovating on higher protocol levels. Each layer below the layer of innovation gradually become more ossified.

\section{Influence of Ethereum}
With the invention of Ethereum, (by Vitalik Buterin. et al), the notion of smart contracts has taken on a more general purpose, compared to the "scripting" in Bitcoin's implementation. The approach of developing a virtual machine, and scripting language for a general purpose protocol has served as a catalyst for the belief of "best protocol wins". Exposing a turing complete scripting language comes with fundamental security concerns. This has led development teams to attempt to build "any" blockchain application on top of the system, and other development teams to build competing general purpose protocols. This work proposes moving away from the general purpose protocol approach, towards dedicated, purpose-built blockchains for a given decentralized solution. 

Also purposed as a transaction-based state machine, similar to Bitcoin, Ethereum attempts to use the transaction model to invoke a state transfer function, while keeping a record of these transitions; Ethereum defines a state transition function to be \(  \Upsilon \), transitioned by way of a transaction, \(T\). Since the Ethereum protocol has been in development, the developers' commitment to their model result in constantly proposing upgrades to it's protocol design, and network properties -- instead of changing fundamental design characteristics of the platform itself. This is a testament to the evolution of techniques used in the decentralized space. 

Similar to Bitcoin, but more formalized in the Ethereum work, the subsequent state is defined as,

\begin{equation}
\sigma_{t+1} \equiv \Pi (\boldsymbol{\sigma}, B)
\end{equation}

where \(\Pi\) is the block-level state transition function, of which is a function of the state \(\boldsymbol{\sigma}\), and the block \(B\) \cite{yellowp}.

\subsection{The Halting Problem}
Turing complete systems succumb to the problem of not being able to determine whether or not a command executed by a turing machine will return a result, or run forever. Concretely, the halting problem is undecidable over turing machines. To circumvent this, the Ethereum project invented this notion of "Gas", an economic property of the system. Around the time of writing, Ethereum users spent an all-time high to send transactions on one day. "Gas" is a measurement of computational "effort", and functuates by demand. These economic principles of the system are decided upon by the protocol developers. Whether a program executes correctly, or throws an error, the submitted of the transaction still has to pay "gas". 

\subsection{Comparing GIT Behavior}

Building from figure \ref{git_ana_both}, by comparing several tokens listed on a cryptocurrency index, GIT behavior provides insight for development activities, and more. Since the majority of blockchain platforms are open sourced, we can use this to analyze past, present, and potentially future development. All measurements are of the time of writing CMC data analysis (June-August 2018).

\begin{figure}[ht]
\centering
	\includegraphics[width=\textwidth]{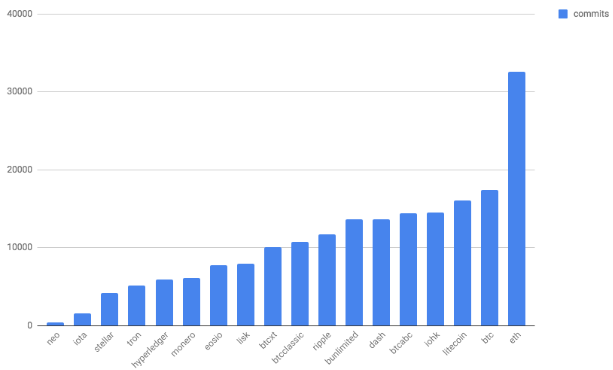}
	\caption{Commit behavior. Five right-most projects are Ethereum, Bitcoin, Litecoin, IOHK (Cardano), and Bitcoin ABC.}
	\label{commits}
\end{figure}

Amongst the top 17 projects, Ethereum's Aleth project maintains the highest amount of commits. Not surprisingly, as many projects under Ethereum's general-purpose approach are actively under development.

\begin{figure}[ht]
\centering
	\includegraphics[width=\textwidth]{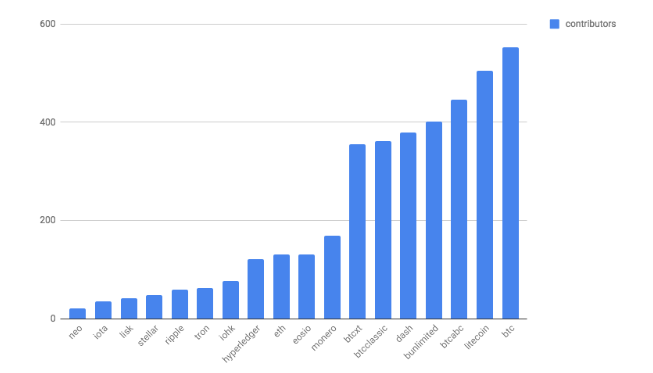}
	\caption{Contributors behavior. Five right-most projects are Bitcoin, Litecoin, Bitcoin ABC, Bitcoin Unlimited, and Dash}
	\label{contributors}
\end{figure}

In figure \ref{contributors}, we show the number of contributors for the repository used in this analysis (most used \& most active). This is interesting, because several of the top projects in this category are Bitcoin Variants -- thus inheriting the number of Contributors during their initial repository forking.

\begin{figure}[ht]
\centering
	\includegraphics[width=\textwidth]{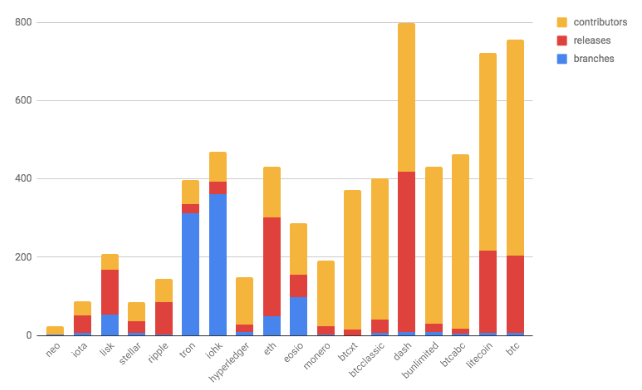}
	\caption{Contributors, Releases, and Branches behavior. Five right-most projects are Bitcoin, Litecoin, Bitcoin ABC, Bitcoin Unlimited, and Dash}
	\label{contrib_releases_branches}
\end{figure}

Figure \ref{contrib_releases_branches} Ethereum is amongst the average within the sample shown, slightly above average on branches, and above average on releases.

\begin{figure}[ht]
\centering
	\includegraphics[width=\textwidth]{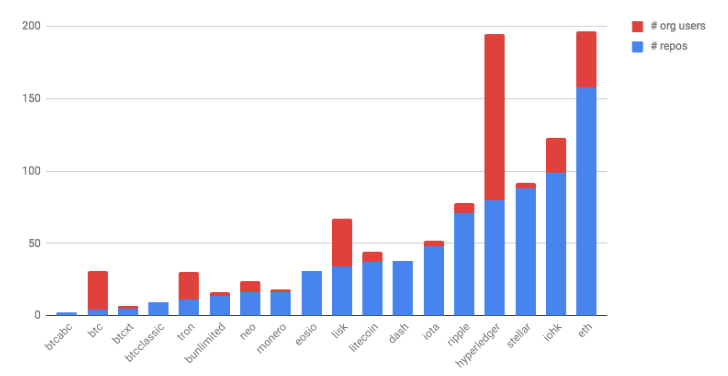}
	\caption{Organizational Users for the project's Github Organization. Five right-most projects are Ethereum, IOHK (Cardano), Stellar, Hyperledger, and Ripple.}
	\label{org_users_repos}
\end{figure}

The insight from figure \ref{org_users_repos} is noticing Ethereum leading the pack of 17 in total number of repositories under the organization account. It is also second regarding the number of organization users. This coincides with collaborative nature of development activity on Ethereum, and the amount of rapid innovation (users creating new repositories). 

The popularity of Ethereum has influenced new protocols to adopt the same methodologies, and design principles. This may be changing as time progresses, but the fact Ethereum has shown developers care about performing activity on a blockchain beyond financial applications.

\chapter{Current Problems}

\section{Challenges}
Showing challenges

\subsection{Decentralization}
A purely decentralized ecosystem would exist in a state where every node within the ecosystem has equal priviledges in a decision, or an equal amount of resources within the system, etc. Whether that decision be to validate a piece of information, or to decide what is a source of truth, it still remains a decision to be made in the system.

Bitcoin's blockchain protocol introduces a tradeoff among consensus speed, bandwidth, and security. By improving the former two, one introduces an increased number of forks, leading to a loss of the mining power that secures the system and to reduced fairness \cite{scalable_blockchain_tech_report}.

This three-way tradeoff, however, is not inherent in decentralized cryptocurrencies. The GHOST protocol \cite{secure_high_rate_transaction} of Sompolinsky et al. as well as Lewenberg et al. \cite{inclusive_block} demonstrate that fairness and mining power utilization can be improved by changing the chain selection rule, in particular, by being inclusive to forks outside the main chain as well. In more recent work, Bitcoin-NG \cite{secure_high_rate_transaction} demonstrates that the inherent tradeoffs in Bitcoin can be eliminated with an alternative blockchain protocol, offering a consensus delay and bandwidth limited only by the Network Plane.

\subsection{Scalability}
Scalability typically refers to a blockchain network not being required to hold knowledge every "transaction" in a system. Currently, many blockchain consensus protocols (eg. Bitcoin, Ethereum, Ripple, Tendermint) have a challenging limitation: every fully participating node in the network must process every transaction. Recall that blockchains possess the quality of decentralization,which means every node on the network processes the same source of "truth".
 
While a decentralized consensus mechanism offers some critical benefits, such as fault tolerance, a stronger guarantee of security, political neutrality, and authenticity, it comes at the cost of scalability today. The amount of data a blockchain can process can never exceed the capacity of a single node participating in the network -- without introducing serious sacrifices in security guarantees. In fact, a blockchain can become sub-optimal as more nodes are added to its network due to inter-node latency that logarithmically increases with every additional node.

\subsection{Overhead}
To develop the simplest local application on the Ethereum platform, for instance, requires us to download a "wallet" to our machine. Otherwise, we have to pay money to test our application, or receive the currency from a testnet.

\begin{quotation}
Would developers be willing to accept any decrease in centralization if it allowed them to not only develop faster, but also fully deploy faster? 
 \end{quotation} 

Developers can create a private network from a genesis block, but they must still abide by the rules implemented by the developers of Ethereum. Developers can also alter the Ethereum Core code base, but this requires much errort. In addition to develop a custom use case using an Ethereum smart contract, developers must learn the Solidity programming language, This passes additional effort onto the developer, and not the system. This is a ramification driven by the assumption that one should develop the "best", general purpose protocol.

\subsection{Capabilities}
With current blockchain implementations, their may be a limit on the size of data a developer is allowed to "store", a specific hash function used in the implementation, or a specific fee structure imposed upon the users. This is caused by developers not participating in the design of these cryptoeconomic rules. This article proposes a system to enable these decisions to be made by developers, upon the developers choosing to do so. Design decisions yielded to the developer are not a centric concern for the developer of blockchain ecosystems today. 

Providing a proprietary turing complete programming language may not be the answer, as it addresses the customization concern at a much higher level, and specific to smart-contracts -- not the cryptoeconomic system as a whole. Providing a turing complete language does solve the concern of customization for developers, but along with the implicit challenges associated with it. The overall concept of a blockchain, consensus algorithms, and cryptoeconomics proves to be a difficult topic for many developers. Adding additional, arbitrary complexity only complicates the development process.   

\subsection{Evolution}
Today, blockchain ecosystems succumb to the requirement of implement new "features" to their design, and some need to "invent" ideas of how to handle problems as the design evolves. 

\subsection{Simulation Tools}
Many blockchain systems today lack meaningful, and powerful developer tools for simulation. In Ethereum for example, one must download local tools onto ones machine for any simulation. This project also aims to explore exposing useful developer tools for simulation of an arbitrary economic design.

\subsection{Startup Time}
There exists not a platform for quick blockchain simultation, debugging, and production setup/deployment. The development process for may platforms today require many steps of preparation, and education. This project aims to enable a familiar workflow for developers to test decentralized ideas.

\subsection{Applicability}
Platforms like Ethereum, having a turing complete language allows developers to model any problem, assuming they are willing to use the rules of the Ethereum platform. This platform enables developer-centered customization capable of re-creating Ethereum, let alone any blockchain architecture. For example, one can expose a programming language to not only create, manage, and deploy a smart contract, but to also deploy an Ethereum clone if one wished to do so.
 
The power behind such a platform allows developers to simply deploy the most naive, simple Blockchain economy, or a complexed ecosystem with "smart functionality". The notion of a "smart contract" is arbitrary, and only reflects the idea of a logic-based, binding agreement between parties that exist on a blockchain. This idea can be called, a "smart contract", "smart object", or a "bubble gum easter egg", the notion is the same.

\subsection{Developer Friendliness}
In order to become a node within the Bitcoin, or Ethereum networks, one must deploy the platform's core source code onto a computer wishing to join the network. This can be an arduous task for a lesser experienced software developer. Developers have the option to use the core source code of one of these blockchain platforms, and deploy a private network, starting from the genesis block. Developers can also alter the genesis block from which to start a new network.

In order for developers to alter any property of the implemented protocol, such as the consensus algorithm used, the hashing algorithm, or the size of each block, the developer has to alter the core source code of the protocol. This can lead to unexpected behavior of the protocol, but is a necessary step for developers seeking a more customized protocol. It is a difficult challenge to shoulder the load of designing the perfect protocol, or the most general purpose. However platforms today inherently commit to their protocol design, and search for ways to apply their protocol to solve problems. It is our belief that it is more feasible to design a \textbf{protocol designing process} for developers to develop, test, deploy, and maintain customized blockchain protocols, at scale, and with the option of abstracting away all operational details of the developer's protocol design/implementation.

\subsection{Turing Completeness}
A computer is Turing complete if it can solve any problem that a Turing machine can, given an appropriate algorithm and the necessary time and space/memory. When applied to a programming language, this phrase means that it can fully exploit the capabilities of a Turing complete computer. One of the main appeals of platforms such as Ethereum is the availability of a turing complete programming language. However, if something is turing complete, as any other benefit, there arises consequences. This consequence can manifest itself in the form of security concerns. 

For a language such as Bitcoin script, being that it is turing incomplete, there are theoretically infinite problems such that a turing incomplete language cannot solve. By the same theorem, there is a finite set of problems such a language can solve. This theoretical constraint, though true, can be over exaggerated, and can depend purely on the implementation. Meaning, the cardinality of the finite set of problems a turing incomplete program can solve may still very high, from the perspective of practical implementation. 

Since the number of computations required in a turing complete language is potentially unbounded above, a malicious miner could always include a block with a transaction purposed for rewarding themselves with fees, in an loop. This could also happen recursively in a non-turing complete language.

\begin{figure*}[ht]
\centering
	\includegraphics[width=0.8\textwidth]{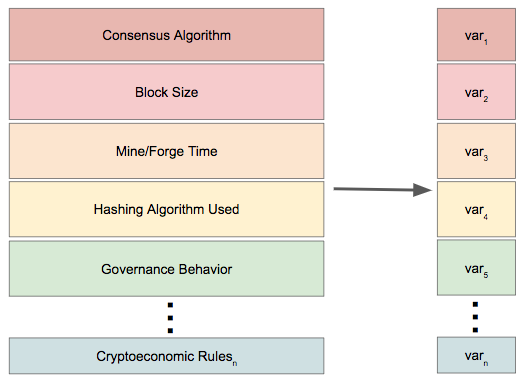}
	\caption{Regarding cryptoeconomic protocol design, there are several design decisions to be made, of which all influence protocol behavior}
	\label{reasons_modularity}
\end{figure*}

\subsection{Security}
Security can refers to being able to withstand an attack by \(greater than\%1\) of the network. Systems such as Bitcoin assume at least \%51 of the network is "honest"; "honest" referring to not malicious, and following the rules -- this is known as "honest" majority.

\subsection{Capacity Increase}
To increase capacity, it isn't a case of using single methods to solve the problem. Developers are choosing many approaches, and combining them. The limitations of the amount of transactions that can be processed is bounded by a few factors. One of course being morse law, and the speed of light. More practically, the limitations include

\begin{enumerate}
	\item Overlay Networks
	\item Blocksize
	\item Message Propagation
	\item Optimizations in Validation
	\item Optimizations in Processing
\end{enumerate}

\subsection{Methods}
Overlay networks allow developers to create a customize protocol layer on top of the main protocol. This is being used in many protocol implementations external to Bitcoin, and is not foreign for Bitcoin itself. The default block size for Bitcoin was 1mb, and have had many proposals to increase it, all with different reasoning. The block size limits the amount of fixed size transactions that can be inserted into it.

\section{Reasons for Modularity}
One of the reasons so many protocol frameworks exist today is the amount of permutations among the different protocol properties. Illustrated in figure \ref{reasons_modularity}, protocols may differ by the hashing algorithms used, to the consensus properties exhibited, even to the size of each block within the chain.  This permutation set of potential protocol behaviors/properties is one of the principal justifications for developing a framework of which allows developers to choose the behaviors that maximally benefit their decentralized use case. Figure \ref{reasons_modularity} shows how protocol governance, block creation time, hashing algorithm, block size, consensus mechanism, and other protocol properties are treated as variables within the system -- among others.

Whenever a team of protocol developers decide on protocol behaviors without taking into account the potential freedom requirements of other developers, it leads to centralized, and difficult-to-alter design decisions being finalized.

\section{Developer Privilege Asymmetry}
There exists an imbalanced set of privileges between core protocol developers, and any other developer accessing the open sourced project. The core developer(s) of the protocol possess the following privileges:

\begin{enumerate}
	\item Decide cryptoeconomic rules of the network
	\item Decide on any fundraising efforts
	\item Decide development decisions
\end{enumerate}

Comparatively, any "external" developer(s) (a developer not existing within the core development team) possesses the following privileges:

\begin{enumerate}
	\item anyone/some public can mine/forge
	\item Anyone can create a transaction/smart contract
	\item Anyone can download the protocol's source code (protocol transparency)
	\item Anyone can view activity
	\item Anyone can propose changes
	\item Anyone can begin a new protocol from original source (usually a fork)
\end{enumerate}

\section{Code Complexity}

The Curse of Code Complexity (\textbf{CoCC}) claims that higher skill-levels are required by any developer looking to take advantage of a complicated, but open-sourced code base. OSS (Open Sourced Software) is built around transparency of code. The more complexed the code base is for a project, the higher the knowledge requirement is for any developer seeking to make changes to it. Blockchain protocols already introduce complexed ideas, and concepts for many developers to deeply comprehend at the time of writing. Adding to that existing complexity is not conducive to developers looking to experiment with cryptoeconomic development as the learning curve can become steeper than necessary.

To develop a customizable cryptoeconomic protocol from an existing project isn't ideal for the average developer for the following reasons:
	
\begin{enumerate}
	\item Must integrate all protocol configurations into that existing protocol
	\item For most, if not all protocols, if you alter the protocol significantly, it will no longer be compatible with the network (forking)
\end{enumerate}	
	
Many projects aim to be the "glue" binding public blockchains, but interoperability, does not guarantee better developer experience. This project aims to lessen the learning curve, and barrier-to-development for the average developer interested in decentralized cryptoeconomic design.

\chapter{The Discipline of Mechanism Design}

\section{Introduction}
Using mechanism design, we can work backwards from a socially acceptable outcome amongst participants, to a set of fundamental behaviors performed by the agents within a game theoretic framework. Mechanism design plays a central role in designing cryptoeconomies. We define behaviors we would like network nodes to perform. Using these desired behaviors, we construct incentives for the purpose of promoting "honest", or "truthful" behavior.

A mechanism involves a finite set of participants, and a set of social decisions to be made. Voting examples are common when explaining Mechanism Design, so consider a group of voters wishing to vote upon a set of candidates; the candidates will be chosen by society as a whole. Each participant's information, usually private, will be referred to as a signal, or the participants "type". The participants report their types to the mechanism, which represents their preference regarding the candidates. For example, participant 1 has a preference for candidate A, opposed to candidate B.

The participant's type can also represent others type of private information to the mechanism. A probabilistic common prior can also exist over the distribution of participant types. Choosing the "best" social decision is dependent on the participant types. We define a decision function mapping types to social decisions.The utility of a participant is a function from their reported type, which may not be reported truthfully, and the output of the decision function.

A social choice function maps reported types to actual outcomes, and this function can possess non-monetary, and monetary aspects. The entity constructing a mechanism has direct influence on the choice of the mechanism, but does not have direct influence regarding the participants, or their reported \& true types. These components make up the environment within which the mechanism works.

We assume an agent knows only their own parameters, but not those of other participants. The designer of the mechanism knows only the environment space, \(\Theta\), and the goal function, \(F\), that is, the class of environments for which a mechanism is to be design and the criterion of desirability \cite{DEM}. Other assumptions can be made as the field of mechanism design is large, and malleable.

In formal notation, we write

\begin{equation}
	F : \Theta \rightarrow Z
\end{equation}

where \(Z\) is the outcome space. In economic models, the outcome space is usually a vector space, but we can take the space of outcomes to be \(Z \in \mathbb{R}\).

A mechanism \(\boldsymbol{\pi}\) can also be viewed as a process for message exchange. In equilibrium form, consists of three component,

\begin{enumerate}
	\item message space, \(M\)
	\item equilibrium message correspondence, \(\mu : \Theta \rightarrow M\)
	\item outcome function \(h\), \(h : M \rightarrow Z\)
\end{enumerate}

Let \(\boldsymbol{\pi} = (M, \mu, h)\). The message space, \(M\) consists of messages available for exchange. We take the message space to be of finite dimensions in Euclidean space. The group equilibrium message correspondence, \(\mu\), holds a relation between an environment, \(\theta\), and the set of messages, \(\mu(\theta)\), that are equilibrium or stationary messages for all agents. These are messages each participant deems as acceptable when the environment is \(\theta\).

The outcome function maps messages into outcomes. Thus, the mechanism \(\boldsymbol{\pi} = (M, \mu, h)\) when operated in an environment \(\theta\) leads to the outcomes \(h(\mu(\theta))\) in \(Z\). The mechanism can be deterministic, and reliably output the same decision and payout, or can be probabilistic/stochastic according to some criteria.

The goal of mechanism design is to generate a mechanism that incentivizes rational participants to perform specific behavior, based upon their private information, thus leading to socially desirable outcomes. A mechanism is said to "implement" a social choice function if, in equilibrium, the mapping from types to outcomes is the same as the mapping that is to be chosen by the social choice function. 

Using several aspects of mechanism design, we can incentive participants in desirable ways, and construct games within which users have no incentive to deviate from behaviors we wish for them to perform. There are several methods to impose dominant strategies, where participants have no reason to behave unfavorably.

\subsubsection{The Revelation Principle}
\indent
One of the foundations of mechanism design is the Revelation Principle. It states that any social choice function that can be implemented by any arbitrary mechanism, can also be implemented by a truthful, direct-revelation mechanism with the same equilibrium outcome. A direct-revelation mechanism is where participants declare their types to the mechanism, resulting in a set of transfers, and a decision. This type of mechanism is "truthful" if participants reporting their true preferences is a dominant strategy for the participants. Such mechanisms are also referred to as truthful, incentive compatible (IC), or strategy-proof.

\subsubsection{Constrained Optimization in Mechanism Design}
\indent

How does one wishing to construct a mechanism ensure the creation of a "good mechanism". Much literature has been written regarding systematic methods of producing mechanisms. The process of creating a mechanism is analogous to solving a constrained optimization problem. The goal being trying to maximize an objective function, under a set of constraints. There are a plethora of constraints used in designing mechanisms, and their coverage if our of scope for this paper. However, here are just a few,

\begin{enumerate}
	\item Incentive Compatible
	\item individual rationality - no agent loses by participating in the mechanism
	\item Budget Balancing and Weak Budget Balancing
\end{enumerate}

A challenge that arises from mechanism design is that some constraints are often impossible to simultaneously satisfy under incentive compatibility; several impossibility theorems have been proven to demonstrate this. Upon imposing constraints, several mechanisms are usually available to choose among. Overall, mechanism design enables us to make assumptions about the behavior of participants within the environment. The weaker behavioral assumptions we impose, the more plausible our theoretical predictions on what happens in a system.

\subsection{Vickery Auction}

In an auction, for which we'd like to compose a mechanism, If bidders bid truthfully, then the auction maximizes \texttt{Social Surplus}, which by definition is just the sum of the values of the winners. In single auction, there is one winner, so its just the value of the winner. Which can be Social surplus can be defined as,

\begin{equation}
\texttt{Social Surplus} = \Sigma v_i x_i
\end{equation}

where \(v_i\) is the valuation of the bidder, meaning how much they are willing to bid, and \(x_i\) is an indicator of whether or not participant \(i\) has won. \(x_i\) is 1 for the winner, and 0 for everyone else.

\begin{equation}
	x_i \equiv
	\begin{cases}
	 1 & \texttt{winner}(x_i) = true \\
	 0 & \texttt{otherwise}
	\end{cases}
\end{equation}

This concretely ensures the winner only pays the amount they bid.

\subsection{Incentive Compatibility}
Due to the Revelation principle, we can strictly represent Incentive Compatible (IC) mechanisms as defined through game theory-based mechanism design. 

\subsubsection{Notation}
We write a mechanism as, also as \(\boldsymbol{\pi}\), depending on the text.

\begin{equation}
M = \boldsymbol{\pi} = (f, p_1, \ldots, p_n)
\end{equation}

we define a social choice function as,

\begin{equation}
f: V_1 \times \ldots \times V_n \rightarrow A
\end{equation}

Which maps from the set of preferences, to outcomes. Which is the cartesian product among all individual preferences. Therefore a subset contains an element from each of the participant preferences, each being a ranking.  Preferences are rankings in this sense, but can be any abstraction of private information. This is private information held by a participant in the economy in this context. 

In Bitcoin, private information takes several forms. From a miners perspective, upon successful mining, its newly found "Nonce" is private information at that time. Once submitted to the network, a mechanism then possesses information regarding whether the submission warrants the granting of a mining reward or not. 

we defined Payment functions as,

\begin{equation}
p_i : V_1 \times \ldots \times V_n \rightarrow \mathbb{R}
\end{equation}

which takes the preferences of the participants, and maps to a real number. For demonstration we restrict this to a real valued scalar, but this can take higher dimensions.

We also define the valuation function, \(v_i \in V_i \) as,

\begin{equation}
v_i : A \rightarrow \mathbb{R}
\end{equation}

which maps an outcome to a real value. A mechanism, \(M\), is Incentive Compatible if and only if, for all \(i\), and \(v_{-i}\) (other participants):

\begin{enumerate}
\item Payment \(p_i\) does not depend on \(v_i\), but only on the chosen social choice, \(a \in A\)
\item \(M\) optimizes for each participant
\end{enumerate}

If a participant has no incentive to misreport their type, or preference, their incentives are compatible with the mechanisms goal. Since \(p_i\) does not depend on \(v_i\), for each outcome, there exists a payment, such that for every valuation function, \(f(v_i,v_{-i})\),

\begin{equation}
\forall a \in A : \exists p_a \in \mathbb{R} : \forall v_i \in V_i : f (v_i, v_{-i}) 
\end{equation}

We can then define \(p_a,\),

\begin{equation}
f (v_i, v_{-i}) = a \rightarrow p_i (v_i,v_{-i}) = p_a 
\end{equation}

If \(M\) optimizes for each player, then we define

\begin{equation}
f(v_i, v_{-i}) \in \operatorname*{argmax}_{  a \in f(v_{-i})  } (v_i(a) - p_a)
\end{equation}

By \(M\) optimizing for each player, \(i\), the mechanism further ensure no incentive to misreport preferences, or types. However, this does not mean misreporting type is uniquely the worst choice, but it does imply there will be scenarios within which misreporting type is economically suboptimal.

\subsubsection{IC Definition}
Dominant strategy incentive compatible means that a participant reporting their true type, or bidder submitting their true bid, is a dominant strategy. If a participant provides their true type, they are guaranteed non negative utility by the mechanism used.

A mechanism \((f, p_1, \ldots , p_n)\) is called IC if for every player, \( \forall i \), every \(v_1 \in v_1, \ldots , v_n \in V_n \) and every \(  v_{i}{'} \in V_i \), if we denote \(a = f(v_i,v_{-i})\) and \(a_{}{'} = f(v_{i}{'},v_{-i})\), then \(v_i(a) - p_i(v_i,v_{-i}) \geq v_i(a_{}{'}) - p_i(v_{i}{'},v_{-i})\) \cite{DEM} \cite{AGT} \cite{ICM}.

\subsubsection{Exploiting Mechanisms}
\indent
The proposed system treats Mechanism as first class citizens, and enables developers to create mechanisms, and use them as primitives. Additionally, consensus within the system is constructed using at least one mechanism.

For example a simple Proof-of -Work (PoW)-style mechanism can be designed by implementing a mechanism that accepts a Nonce as input, and returns a boolean decision ("accept" or "reject"). Next we would build a mechanism regarding longest chain wins, and the actions taken after receiving a Nonce from another node, etc. Participants simply must invoke the overall mechanism to retrieve a result from its parts. To implement PoW-style consensus, we experimentally expose other data types, such as a "Puzzle". Puzzles allow developers to create arbitrary computations that accept solutions from participants. For simple PoW, with \(z_L\) being the number of leading zeros required, or "target value", that the resulting solution hash \(s_{hash}\) must be lower than, the puzzle is:

\begin{equation}
	\texttt{substring}(s_{hash}, 0, z_L) =  "0 \ldots 0"
\end{equation}

For every solution submitted, the protocol can check whether or not this Puzzle has been solved by the candidate input. These are examples of the application of mechanism design to the actual design of decentralized economies. The platform discussed supports experimentation with creating mechanisms. Current platform that enable "smart contract" allow developers to already create mechanisms. The proposed system allows developers to design all mechanisms in the system, and not simply an additional mechanism atop several others -- to construct decentralized systems using mechanisms as foundation.

\part{Proposal}

\chapter{Protocol \& Platform}

\section{Mission}

\begin{figure*}[ht]
\centering
	\includegraphics[width=0.9\textwidth]{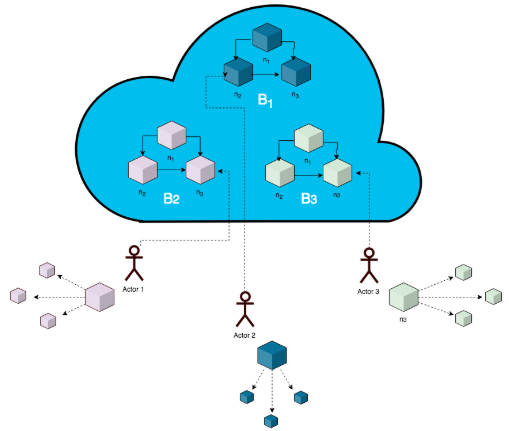}
	\caption{The architecture consisting of network nodes, both local to an actor, and distrubuted, existing with additional, heterogeneous chains operating from the same protocol.}
	\label{kunta_arch}
\end{figure*}

This article proposes a system to enable customization of decentralized protocols, create a cryptoeconomy implementing the designed protocol configuration, usage of that economy's properties as a usable data type in modern, arbitrary programming practices, and an example of a programming language purposed with creating and interacting with the deployed blockchain.

Due to the inherent tribalism regarding the "best protocol" wins zeitgeist of today, the status quo moves more towards generalized, multipurpose, single design protocols. The opposite being a specialized protocol design for every unique solution, given certain common fundamental design factors. The specialized blockchain implementations tend to be built on top of an existing single design protocol, or lacking usefulness outside of the context for which it was designed, without massive refactoring, and redesign.

This effort also emphasizes the need for a process for creation, deployment, and management of any protocol implementation. We believe there is much room for experimentation with decentralized architectures, both purely decentralized, and semi-decentralized; zealots for the former often miss opportunities for innovation along the way.

Conceptually, we wish to formulate an abstract machine purposed with deterministically producing a dedicated cryptoeconomic protocol from a set of properties.

\begin{quotation}
  We define a cryptoeconomic protocol as a mechanism consisting of a goal function, and environment
 \end{quotation}

\subsection{Notation}
As convention, we denote an arbitrary blockchain as \(\boldsymbol{\mathcal{B}}\), 

\begin{equation}
	\boldsymbol{\mathcal{B}} \equiv (B_0, \ldots, B_n   )
\end{equation}

consists of several, sequentially linked blocks, \(B_h\), \(n\) being the number of blocks in a chain, indexed by the height of the block, \(h\). Each \(B\) contains various values relevant to the context of \(B\),

\begin{equation}
T \in \boldsymbol{T}, T \equiv B_{n_{T}} \equiv T(B_n)
\end{equation}

the transactions, in the conventional sense, for a given block can be denoted by \(B_{n_{T}}\) or \(T(B_n)\), but represents all transactions in a given block \(B\). The block header, \(B_h\) can contain various types of information, but at least has to contain all a presentation of all \(T\) belonging to it, usually representing by hashing the set of transactions, or a Merkle Tree (CITE) root hash of all \(T\) in \(B\).

\begin{equation}
\forall B, B_h \equiv (\ldots \texttt{H}(T_0, T_1, \ldots))
\end{equation}

\(B\) is generated through computational means, and finalized through a block-finalization state transition function is defined as,

\begin{equation}
\Pi (\boldsymbol{\sigma}, B) \equiv \Omega(B, \boldsymbol{\Upsilon}(\Upsilon(\boldsymbol{\sigma},T_0), T_1) \ldots )
\end{equation}

with \(\boldsymbol{\Upsilon}\) being a protocol's state transition function, and \(\Upsilon\) being the transaction-level state transition function. Upon finalization of \(B_h\), \(\Omega\), the protocol can choose to reward an account -- either through nomination (privilege reward), or by mining (value reward), depending on the consensus properties of \(\boldsymbol{\mathcal{B}}\) designated by the implementing developer. \(\Pi\) denotes the overall state transition function for \(B\).

\subsection{Blockchain Types}
We expose a set of customization configurations for developers to choose between. This set is intended to be grown over time, as the academic literature grows on blockchain protocol design. Among these configuration options, we define two variants of blockchain types, \(\boldsymbol{\mathcal{B}}_{type}\), intended to represent fundamental blockchain design concepts:

\begin{enumerate}
 \item Type
	\begin{enumerate}
 		\item Unspent Transaction Output - \(\boldsymbol{\mathcal{B}}_{\mathcal{U}}\)
		\item Account Based - \(\boldsymbol{\mathcal{B}}_{\mathcal{A}}\)
	\end{enumerate}
\end{enumerate}

A goal is not to restrict developers to one Blockchain type, or try to develop a protocol attempting to generalize them all, but to give developers the choice of \(\boldsymbol{\mathcal{B}}_{type}\), and grow the amount of choices as the Blockchain Ecosystem grows, and exposing the common factors of \(\boldsymbol{\mathcal{B}}_{type}\).

\subsection{Object Gender}
Consulting the Bitcoin implementation, the notion of inputs and outputs is a central theme to its transaction model; we can abstract away the idea of unspent transaction outputs (UTXOs). Bitcoin transactions take on the form of an output, or an input. Specifically to the Pay-to-Public-Key-Hash (P2PKH) model, outputs contain Public Key hashes of a targeted account, \(a\), denoted as \(\boldsymbol{\texttt{H}}(K_{public}[a])\), from which to prove an account can rightfully redeem a unit owed to it. 

Inputs contain an account's signature, \(a_{signature}\), consisting of private key encryption of the transaction, followed by a hash of the encrypted result, and their full public key, \(K_{public}[a]\). Bitcoin Script (CITE) then hashes the public key given to it by the redeeming account, \(\boldsymbol{\texttt{H}}(K_{public}[a_r])\), to verify the account's entitlement to the unit. Abstracting away the concept of an input, and an output further, we can apply it to other protocol types. For this, we use Object Gender, \(G\), as used, in subscripts, to classify a "transaction" in general. Some blockchain protocols purposely avoid gender in their design. For example, Ethereum's Ommers nomenclature is purposely derived from the gender neutral concept of a sibling, and "Ommer".

Differences in \(G\) refer to the object/transaction "accepting" invocation, or the object "performing" the invocation. This configuration enables the property of sequence regarding transactions, and can be "self-illustrated" by its nomenclature. This implicitly enables a notion of time to be captured, and inherently forms a "history" of objects occurring in \(\boldsymbol{\mathcal{B}}\). For simple transactions, their can exist only one combination of a female-to-male pair. However, this can generalize into a one-to-many relationship between females and males (similar to Bitcoin's MultiSig), and even more complexed mechanisms. More on this is covered in section \ref{chaintypes}. 

This property can hold for a simple Asset Transfer blockchain, \(\boldsymbol{\mathcal{B}}_{\mathcal{U}}\) with no denominations, and a static value for each \(T\), as the transactions are simplified versions of a UTXO transaction, with only a 1-to-1 correspondence of males and females. With \(\boldsymbol{\mathcal{B}}_{\mathcal{T}}\), each transaction contains only one female, for inserting an asset in \(\boldsymbol{\mathcal{B}}\), and one male for invoking an asset in \(\boldsymbol{\mathcal{B}}\), for transfer, or validation, etc. The male invocation can result in a myriad of operations occurring, all defined by the  \(\boldsymbol{\mathcal{B}}\) specification created, and the virtual machine's execution upon the invoking transaction. This is covered more in the section on the Virtual Machine implementation.

In an account-based chain, \(\boldsymbol{\mathcal{B}}_{\mathcal{A}}\), such as Ethereum, differences in \(G\) exists in the code execution processes. Upon a contract code creation function, \(\Lambda\), being executed, the transaction used is here referred to as Female, \(f\), one that accepts invocation. The addressing in Ethereum specifies an account's address, \([a]\), of which can represent an account to be referenced within the network, a 20-byte hash, \(\mathbb{B}_{20}\), and an empty byte set, \(\mathbb{B}_{0}\) for a "smart contract"; of which theoretically have the same capabilities in Ethereum, by design. Instead, we simply use \(\boldsymbol{\texttt{H}}(T)\) to refer to "targeting" a transaction, \(\texttt{to}(T)\). Generally, \(T_m\) targets \(T_f\), but we can link transactions for deeper invocation.

With Ethereum, the contract creation code, \(c\), executes one time only, and returns the body of the contracts code to be stored, and executed every time the contract receives a message in the network, or is invoked from a future transaction, referred to as a "message call". More on \(\Lambda\) is covered in section \ref{accbase}.

The message calls, \(\Theta\), to an existing contract, post creation, can be viewed as Male transactions, \(m\). These transactions execute the \(f\) transaction it is referring to, and uses the data passed by \(m\) as the parameters, \(m_p\), to the \(f\) code being executed. More on \(\Theta\) is covered in section \ref{accbase}. Parameters in Bitcoin are simply concatenated to the output in the transaction, and the concatenated bitcoin script code is executed on that.

Drawing analogy from electrical, and mechanical engineering, \(G\) can be viewed as similar to "headers" at the wire tip of certain electrical components. \(m\) header pins are extruding, for the purpose of inserting into \(f\) header sockets. Due to differing nomenclature between protocols regarding objects or transactions accepting, and triggering invocation, referring to \(G\) allows us to completely abstract away object and transaction types, regardless of protocol today; this may change over time.

\subsection{Value}
In cryptoeconomics, one must incentivize computation, or the securing of the chain through time; an agreed upon method is required for the transference of value within the network, whatever is the definition of value for that network. The answer to this has traditionally been to formalize an arbitrary currency, along with a set of rules to govern its usage within the network.

We purposely abstract away the notion of "value" for the purpose of generalizing its definition. Using this, we can construct an instance within which value is a currency, or an asset, or privileges, and so on into the imagination of the developer. 

\begin{quotation}
  \(\boldsymbol{Value}\) in a given cryptoeconomy is defined by the implementing developer. Thus we do not limit the game theoretic applications of the protocol.
 \end{quotation}

\chapter{Protocol Customization}

\begin{figure*}[ht]
\centering
	\includegraphics[width=0.8\textwidth]{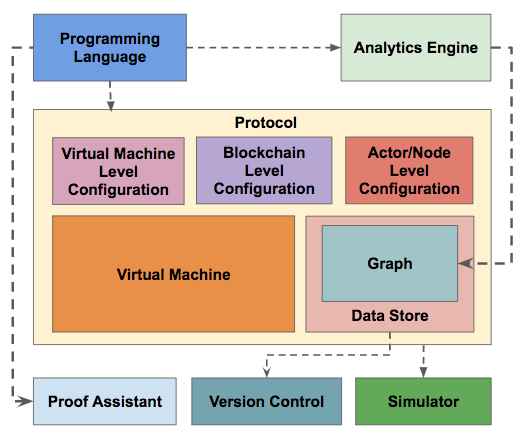}
	\caption{The technology stack used}
	\label{k_architecture}
\end{figure*}

\section{Configuration Layers}
As a lesson taken from the computational simulation world, we can construct hyperparameters around any blockchain protocol, assuming the founding developers allow it, without breaking "honesty" in the network. We do not separate these hyperparameters into mutually exclusive roles, but categorize them into three layers of customization for design purposes. Though overlap can exist between them, this serves as a heuristic towards designing the protocol design process. This is not proposed as the correct way to do this, but serves as a starting iteration on the approach of modularizing protocol design aspects into a configurable set of state machines.

The three layers of customization we focus on are:

\begin{enumerate}
 \item Actor/Node  - \(\texttt{A}\)
 \begin{enumerate}
		\item Consensus
		\item Mechanisms
	\end{enumerate}
 \item Blockchain - \(\texttt{B}\)
 \begin{enumerate}
		\item Type
		\item Roots (set)
		\begin{enumerate}
			\item Aspects
		\end{enumerate}
		\item Hashing Algorithm
	\end{enumerate}
 \item Virtual Machine - \(\texttt{V}\)
 \begin{enumerate}
		\item Operation Codes 
		\item Compute Constraint
	\end{enumerate}
\end{enumerate}

Therefore any chain created must be characterized by the cartesian product,

\begin{equation}
	\texttt{Create} : \texttt{A} \times \texttt{B} \times \texttt{V} \rightarrow \boldsymbol{\mathcal{B}}
\end{equation}

The overall goal of the protocol is to satisfy the relation,

\begin{equation}
\forall \boldsymbol{\mathcal{B}}, \boldsymbol{\mathcal{B}} \subset \texttt{A} \times \texttt{B} \times \texttt{V}
\end{equation}

\subsection{Actor}
For actor specific customization, developers can choose the type of choice model (uncoordinated/coordinated) the network uses. There are also properties to customize that may lie between the Actor and Blockchain configuration layer. For example, choosing the consensus to use isn't exclusively relating to either one, but affect both. We believe consensus will continue to be a widely experimental property of any blockchain network. For this reason, this project lends itself towards supporting continued experimentation with newer, and emerging approaches.

\subsection{Chain Type} \label{chaintypes}
We initially expose two fundamental chain types, both with the same level of customization, and the same level of code execution capabilities. We define both, and eventually many, instead of one, as there still exists schools of thought around the "better" approach on chain type.

\subsubsection{Unspent Transaction Output}
Inspired by the original Bitcoin, an unspent transaction output (UXTO) chain requires each \(T\) to be spent, must contain a male segment. Upon a male invoking the female segment of \(T\), it creates at least one, or more female segments, of which the sum of female segments equals the value, \(\boldsymbol{\texttt{V}}\) stored in the female segment being invoked. Ignoring fees, upon spending of a transaction, the following inequality holds,

\begin{equation}
	 \sum \boldsymbol{\texttt{V}}(f) \geq \sum \boldsymbol{\texttt{V}}(m)
\end{equation} 

By definition in a UTXO style blockchain, since each output requires an input to be spent, the maximum amount of \(T_m\) allowed to invoke the entire transaction is the total number of \(T_f\), 

A simple asset transfer type is a UTXO type with static value per transaction, and a 1-to-1 transaction relationship, unless supporting multi-transfer. Every transaction is either a female, or male, and corresponds to one of the opposite gender. This can be used for simple asset exchange.

\subsubsection{Account Based}
An account-based chain type maintains a world state, \( \boldsymbol{\sigma} \), representing the state \(\texttt{tree}\), of which is implemented as a key-value store of accounts, \(a\), of which each corresponding to a specific account, and its world state, \( \sigma [a] \). This world state is usually stored locally to the node -- contrary to being stored "on chain" as is popular belief.

For efficiency, Ethereum tracks a state tree (specifically, a trie data structure), holding all account states in the network; contract code on the chain are also included in the state, as they maintain the same capabilities as user accounts. 

\begin{equation}
L_{I}\big( (k, v) \big) \equiv \big(\texttt{KEC}(k), \texttt{RLP}(v)\big)
\end{equation}

They define a "collapse function" around \(L_{I}\) for the set of key pairs in this tree. This is the fundamental representation held in the Ethereum network at the time of writing.
Developers can choose to deploy a private chain, and not inherent meaningless world states from the network. However, as mentioned before, this requires abiding by the implemented protocol, and allows no room for flexibility without breaking "honesty", or requiring reverse engineering of the protocol itself.

Developers should be able to define their own world states to exist in the network, without being required to inherent every other world state by other users irrelevant to their ecosystem, and without unnecessary developer friction hindering new developers.

\subsection{VM}

The role of a virtual machine is to simulate the execution behavior of a computer, within a specific context. The majority of protocols today implement their own, custom virtual machine (if any), for the purpose of powering a proprietary scripting paradigm upon which their protocol runs. 

These protocols fall victim to the same single-design mentality as the core implementation requires, with little-to-no level of customization to the developer community, without causing hard forks in the network. Developers can customize the opcodes being executed within the virtual machine, although we aim to expose an expressive VM with which to begin. 

Additional security concerns exist this, but are navigable through best practices; this proposal can still exist in its entirety without the capability to customize the underlying virtual machine.

\chapter{Abstracting Block Roots}

\begin{figure}[ht]
\centering
	\includegraphics[width=0.5\textwidth]{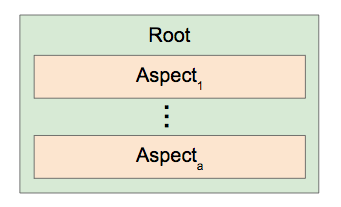}
	\caption{Aspects are dynamic variables used in the system, and they are associated with a Root}
	\label{aspect_in_root}
\end{figure}

\section{Root Instances} 
Designs have implemented developer-chosen Merkle Trees (or tries), and Merkle Roots to store families of data, and sequences. The decision of what to use a Merkle Tree for should be left up to the developer, and not taken at face value. For example, keeping record of the "Ommers hash", or hash of the "uncles" within Ethereum's design may make sense for it, but should't be taken as a generalized design choice required for a blockchain powered application. The choice of Merkle trees, and roots to include in a block's header (in the case of merkle roots), and in the body of a block on full nodes (optionally, merkle trees) should be left to the implementing developers, not the platform. 

As stated before, the merkle Root of the transaction merkle tree in a block is hashed, and included in a block. This serves as cryptographic proof of any transaction in that tree. Every other root included in a block (transaction root, state root, receipt root, etc), is included for the same purpose of "proving" it occurred, or existed. Other protocols also use hashing of the root of a tree data structure to "prove" existence of an element within the tree. We abstract away the idea of a Root in a chain header. This can satisfy the requirements for keeping track of State, \(\boldsymbol{\sigma}\), and \(T\).

\begin{quotation}
  The abstraction of a \(\boldsymbol{Root}\) allows us to expressively generalize the information intended for inclusion in any \(\boldsymbol{\mathcal{B}}\).
 \end{quotation}

We define an abstraction of \(T\) that can appropriately represent the objects that invoke change in a state machine. In the case of Bitcoin, a root instance, would be the general transactions, with the root being user-defined as "Transaction". However, we can represent additional examples of arbitrary roots to include in \(B\), such as the storage, receipt, and transaction treis in Ethereum. These are still sets of objects, from which we can execute logic. The way a protocol handles a storage root, is the same as handling a transaction root, but we execute different functions from the instructions represented by its respective root, of which is user-defined, in the specification of their \(\boldsymbol{\mathcal{B}}\).

\subsection{Male}
A male object simply invokes a female object. Male objects "spend" transactions in a Bitcoin case, and in an Ethereum implementation can serve as any execution of a smart contract, after it has been deployed, or "invocation". For deeper invocation, we can define subinvocation by either \(f\) or \(m\) root instances.

\subsection{Female}
From an Ethereum perspective, the smart contract "creation" function serves as a female object. We deem this as female, as it accepts invocation from male objects. In the bitcoin sense, female root instances represent transaction outputs awaiting inputs.

\subsection{Root Sets}

For each block \(B\),  we include a Root Set, \(\mathcal{R}\), defined as a set of hash values, each of which belong to the indexed, developer-defined root. This is designed to prove any root instances belonging to its respective root, in a chain.

We define \(\Gamma\) as the set of Root Instances, and \(\Gamma_r\) is the set of root instances belonging to their respective root, \(r\), of which is included in a block, \(B\), along with the hash of the root node of the merkle tree of \(\Gamma_r\), \(\boldsymbol{\texttt{H}}(\texttt{TreeRoot}({\Gamma_r}))\).

\begin{equation}
\forall \gamma, \gamma \in \Gamma_r , \gamma \equiv \{\gamma_{n}, \gamma_{t},\gamma_{a}, \gamma_{c}, \gamma_{u}, \gamma_{\boldsymbol{\mathcal{A}}} \ldots\}
\end{equation}

Where \(\gamma_{n}\) is the name of the root, \(\gamma_{t}\) is the type of the root, and \(\gamma_{a}\) is the access option on the root, denoting whether the root is private, or public, \(\gamma_{c}\) is the code defined in the root, \(\gamma_{u}\) is the return instructions for the root, and \(\gamma_{\boldsymbol{\mathcal{A}}}\) is the set of Aspects (more on Aspects covered in "Aspects"). 

\begin{equation}
	{ \forall r_k \in \mathcal{R}, \boldsymbol{\texttt{H}}(\Gamma_{r_k}) }
\end{equation}

The set of roots \(\mathcal{R}\), is created by the configuration of \(\boldsymbol{\mathcal{B}}\). The cardinality of the set of roots, \( \vert \mathcal{R} \vert \), is defined as the amount of roots, \(k\), created by the implementing developer. 

\begin{equation}
\forall r \in \mathcal{R}, r = \texttt{H}( \texttt{Tree}_{root}(\Gamma_r))
\end{equation}

Developers are given the options to define Roots as a part of the protocol design process. We can record a history pertaining to these roots, and the behavior of related root instances over time.

Root Instances have properties, of which we can use for several tasks. For example, if \(\gamma_{G} = m\), \(\gamma\) must contain \(\gamma_{p}\), the partner identifier, a hash of the female partner, \( \texttt{H}(p(\gamma)) \) -- can be arbitrary, but an relationship needs to be made from \(m \rightarrow f\).

\begin{equation}
\Upsilon(\boldsymbol{\sigma},\gamma) \equiv  \begin{cases} 
      \texttt{execute}(\gamma) & \gamma_{t} = f \\
      \texttt{search}(\mathcal{G}, \texttt{H}(p(\gamma)))   & \gamma_{t} = m  \\
   \end{cases}
\end{equation}

where \(f\) is a root instance belonging to the female class, and \(m\) belongs to the male class. If \(\gamma_{t}\) is \(f\), we simply execute on \(\gamma\), and if \(\gamma_{t}\) is \(m\) we first search for \(m_{p}\), where \(p\) is partner of \(\gamma\), found by referencing the partner's hash, \(\texttt{H}(p)\); if no \(p\) is found, throw an error. If we find \(m_{p}\), the protocol attempts to prepare the root instance for execution, and allowance into the network.

We can define simple and advanced security properties around this constraint. For example,

\begin{equation}
\forall \gamma_m, \begin{cases} 
      \texttt{error} & p(\gamma) = \varnothing \\
      \texttt{execute}(\gamma_f \cup \gamma_m)   & \texttt{otherwise}  \\
   \end{cases}
\end{equation}

This approach enables a layer of security around the user-defined specification for \( \boldsymbol{\mathcal{B}} \). No \(\gamma\) type will be executed beyond those of which the implementing developer specifies. Any \(\gamma_m\) that does not reference a \(\gamma_f\) will not be executed, and any male that does not "honestly" invoke a \(\gamma_f\) will also not be executed. We can also impose more restrictions if needed, such as:

\begin{enumerate}
	\item if \(f\) already has been paired with another \(m\)
	\item if the parameters of \(m\) are valid to pair with \(f\)
\end{enumerate}

Each \(\gamma\) contains a segment of code to be executed, \(c\), but its results are handled differently. We highlight the different handlers for each chain type:

\begin{enumerate}
	\item UTXO - The result of code execution "spends" a specific female, by pairing a \(f \gamma\), followed by at least one \(m\). If denominating the outputs, multiple females are created, for each intended "Receiver" of the denominations.
	\item Account Based - The code execution result is stored in the \( \boldsymbol{ \sigma } [a]_{s} \) if \(\gamma\) is a \(f\), to be invoked later by a \(m\), and stored as the result of invocation by a \(m\). This enables root instance-level log receipts. This differs from the separate Root for them included in each \(B\), as in Ethereum, but one can easily create another \(r\) specifically for receipts. We make no designation regarding where to store \(r\) information, and \(\gamma\).
\end{enumerate}

\subsection{\(\gamma\) for UTXO}
In Bitcoin, the parameters of \(m\) are the transactions signature, and the receiver's full public key. This is explained in the Bitcoin developer guide as Signature Script. The receiver's full public key is hashed to verify with the public key hash in the previous output. The node also checks the hash, and encryption (signature) by the receiving address to verify entitlement.

\subsection{\(\gamma\) for Account Based} \label{accbase}
Using an Account Based architecture, each \(m\) and \(f\) \(\gamma\) represent some execution, either code creation, \(f\), or code invocation, \(m\).

\subsection{Female}
For \(\gamma_f\), we define a a transition function,

\begin{equation}
(\boldsymbol{\sigma}', z, \boldsymbol{o}) \equiv \Lambda(\boldsymbol{\sigma}, \gamma_f, s, o, v, \mathbf{c})
\end{equation}

where \(\boldsymbol{\sigma}\) is the current state, \(z\) is the virtual machine status code, \(\boldsymbol{o}\) is the log result for \(\gamma_f\), \(s\) is the source of \(\gamma_f\), usually an account, \(o\) is the original creator, mostly \(s\), \(v\) is the value, if defined in the specification of \( \boldsymbol{\mathcal{B}} \), and \(\mathbf{c}\) is the byte array of code in \(\gamma_f\). \(A\) is the substate, existing prior to execution of \(\gamma\). We can add depth by analyzing substate in-between root instance executions.

\subsection{Male}

For \(\gamma_m\), we define an invocation function \(\Theta\)

\begin{equation}
(\boldsymbol{\sigma}', z, \boldsymbol{o}) \equiv \Theta(\boldsymbol{\sigma}, \gamma_m, s, o, r, c, \mathbf{d})
\end{equation}

where \(z\) is the receipt of \(\gamma\), \(\mathbf{o}\) is the result of \(\gamma\), \(s\) is the sender of the root instance, \(o\) is the originating account of the partner \(\gamma_f\) root instance, \(r\) is the receiving/targeted account \(a\), usually the hash of the partnered root instance, \(\boldsymbol{\texttt{H}}(\gamma_f)\), \(c\) is the code in the root instance, and \(\mathbf{d}\) is additional data. Similar to \(\gamma_f\), we can add depth by utilizing substate between root instance executions.

\subsection{Aspects}
Every \(\gamma\) defined in the speciication contains at least one Aspect, \(\gamma_{\boldsymbol{\mathcal{A}}}\). An aspect can be viewed as a variable for any \(\gamma\) belonging to a Root to reference. Aspects can also be referenced by any \(\gamma\) that does not belong to the respective root, but this must be defined in the specification for \( \boldsymbol{\mathcal{B}} \). 

As illustrated in figure \ref{aspect_in_root}, Aspects are a part of a Root. Within the system, developers can make constant variables, and dynamic variables. Theoretically, constant variables can be included in genesis, or by a later Root Instance, but will never change their value. Comparatively, Aspects, being dynamic, exist on chain, and can be updated. Aspects are included in roots for scoping purposes, but one Root, can access another Aspect through invocation. 

\chapter{Consensus Layer}

\section{Consensus}
Today, there exists a myriad of research literature on several consensus algorithms -- from Proof-of-Work, to Proof-of-Stake, to Proof-of-Burn, and more. However, for platform developers to not allow subsequent developers to easily choose which consensus algorithm to use lessens developer experience. Consensus is a property that can traverse the Actor, and Blockchain layers of customization. Consensus can also span the VM layer as well, but can be constrained to the Actor layer. There is no such thing as a universally "better" consensus algorithm, of which provides fuel for arguments regarding which ones are "better".

\subsection{Consensus as Constrained Optimization}
We, along with others, claim solving the problem of reaching consensus is analogous to solving a constrained optimization problem. We can define the problem with the desired state we wish to reach over the variables considered during consensus.

For example, if we want to arrive at simple consensus, among nodes in a network regarding the current time, we can construct several schemes to do, regarding a myriad of variables, in addition to the current. We can attempt to take an average of all the times reported among nodes. We can also add a measure of geometric distance each node is from each other, and so on. We claim the space of possible consensus variables across which we can construct a consensus mechanism is large.

Once we decide upon a set of variables we wish to consider during consensus  rounds, the problem then becomes a constrained optimization problem. Borrowing from an approach taken by the No Free Lunch theorem for static over optimization algorithms, for any pair of algorithms \(a_1\) and \(a_2\) 

\begin{equation}
	\Sigma_f P(d^{y}_{m} \vert f, m, a_1) = \Sigma_f P(d^{y}_{m} \vert f, m, a_1) 
\end{equation}

Based on the No Free Lunch theorem, we apply the support to designing consensus mechanisms,

\begin{theorem}
	Let \( \mathcal{C} \) be a finite set of consensus algorithms, \( \texttt{c} \in \mathcal{C} \), and \(\texttt{c}\) be a consensus algorithm, \(f\) be a function upon which consensus is desired, \(m\) be iteration steps (i.e decision rounds), \(\texttt{l}\) be latency, and \( \texttt{e} \) be agreement error, for any performance measure, \( \Phi(\texttt{l}, \texttt{e}) \), the average over all \(\texttt{f}\) of \(P( d_{m}^{y} \vert \texttt{f}, m, c )\) is independent of \( \texttt{c} \)
\end{theorem}

where \(\displaystyle d_{m}^{y}\) denotes the ordered set of size \(\displaystyle m\) of the cost values (deviation from consensus, time to consensus, etc) \(\displaystyle y\) associated to input values \(\displaystyle x\in X\), \(\displaystyle f:X\rightarrow Y\) is the function being optimized and \(P( d_{m}^{y} \vert \texttt{f}, m, c )\) is the conditional probability of obtaining a given sequence of cost values, from consensus algorithm \( \displaystyle c \), run \( \displaystyle m\) times on function \( \displaystyle f \). If an algorithm performs well on a certain class of problems then it necessarily pays for that with degraded performance on the set of all remaining problems. The space of all samples of size \(m\) is \((\mathcal{X} \times \mathcal{Y})^m \). In this paper, we only apply NFL Theorem 1. NFL loosely that algorithm \(a_1\) must beat \(a_2\) on just as many target functions (and associated datasets) as \(a_2\) beats \(a_1\). In theory, NFL implies that no consensus mechanism can perform optimally over the entire problem space.

\section{Designing Consensus}
We attempt to formally apply principles of mechanism design to designing consensus. In other words, every consensus mechanism consists of one or more mechanisms. Using this approach, we can natively design consensus mechanism from first principles.

\subsection{Using Mechanisms}
We treat mechanisms as the fundamental building block for each consensus algorithm. Using mechanisms, developers can define the properties of the consensus layer.

\chapter{Virtual Machine Layer}

\section{Specialized Virtual Machine}
As done in Bitcoin, Ethereum, and others, smart contracts execute within a dedicated virtual machine. Inside the proposed virtual machine exists customized operational codes (OpCodes). To design arbitrary opcodes in the service of creating a turing complete language introduces more security risk to any implementation. Instead of attempting to create a general purpose blockchain protocol, we propose to extend the capability to create a purpose-built blockchain protocol catering to the use case of the developer, without propagating the risk of implementation errors onto the developer.

\subsection{Operation Codes}
With systems such as Ethereum, the opcodes are embedded in the system. "Bitcoin Script" enables for very limited customized scripting, and Ethereum enables more, being turing complete. However, these opcode design decisions are not to be made by regular users of Ethereum. A developer must fully attempt to fork the Ethereum codebase, and customize on top of it, which can be an arduous task for any open sourced project contributed to by hundreds of developers. 

Instead of building on top of bitcoin, or Ethereum, the research allows for developers to redesign bitcoin, redesign Ethereum, or more generally design an arbitrary Cryptoeconomic network.

\subsection{Code Execution}
Socially, we define a smart object as any instance created by a developer served with making a decision of some sort. Whether the decision be to retain specific information, calculate an arbitrary function, or consult the world, external to the crypto-economic environment in-use (accessing the "wet" world).

\subsubsection{Gender-Based Execution}
Upon a \( f \) Root Instance, \( \gamma_{G} = f \), we define a code creation function as \( \Lambda \), of which computes on its parameters, the state \( \gamma_{\sigma} \), the sender \( \gamma_{s} \), originating account of the female segment \( \gamma_{o} \), and the instructions to be executed once \( \gamma_{i} \), of which is data, arbitrary in length.

When a \(m\) Root Instance is processed, \(\gamma_{G} = m\), we define the code execution function, denoted \( \Xi \), evaluates to a tuple representing the subsequent state computed \( \boldsymbol{\sigma}^{**} \), a substate \(A\), and the body code of the account, \(c\).

\subsection{System VM}
To operate, we must have some basic operations in the VM. These are in addition to the Operation Codes created by the developer. These are operations for basic arithmetic, cryptographic functions, and other core computations supporting the protocol.

\section{Smart Contract Templates}

 \begin{figure*}[ht]
\centering
	\includegraphics[width=0.75\textwidth]{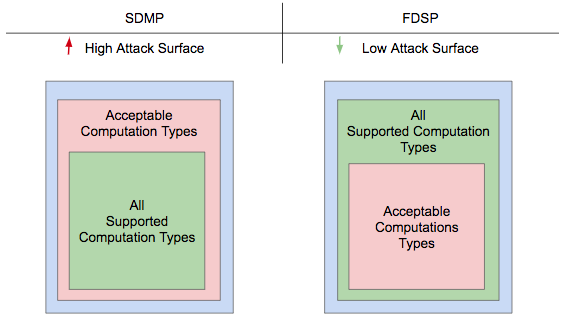}
	\caption{Showing how restricting the smart contract type that can be computed upon naturally lessens the attack surface for a nefarious actor to submit malformed smart contracts}
	\label{contract_security}
\end{figure*}

With Ethereum, the default state of the system is to accept any smart contract submitted by anybody willing to pay the required resources. In the security documentation, there exists ways to restrict transactions from specific parties, or from the public, but these setting are experimental, and not widely used. 

As a consequence, developers must be concerned when deploying a permissionless, public blockchain. This is because nefarious actors can always submit any transaction they can imagine, with little consequences beyond the resources paid to submit the transaction to the network.

\begin{enumerate}
	\item \texttt{\textbf{SDMP}}: Single Design Multi-Purpose 
	\item \texttt{\textbf{FDSP}}: Flexible Design Single Purpose
\end{enumerate}

A \texttt{\textbf{SDMP}} is a protocol that is designed to be multipurpose and general, and the amount of "supported" computation types, and "acceptable" computation types tends towards \(\infty\). These protocol types have one specific design, but users are encouraged to use the protocol "as-is". 

With an \texttt{\textbf{FDSP}} protocol, the "supported" computation types also tends towards \(\infty\), but is bounded above by the implementing developer. This means developers choose what computation are allowed to take place on the network.

This system proposes a concept called "smart contract templating" where implementing developers can specify the structure of all transactions intended to be accepted, and computed upon within the network. 

\begin{quotation}
  By design, \(\boldsymbol{any}\) transaction that does not adhere to the templates set forth in the design of the protocol will not be computed upon by the protocol
 \end{quotation}
 
\begin{equation}
	 \Theta_{result} = 
	\begin{cases}
    	 \Theta(\gamma),	& \text{if } \neg \texttt{ malformed(\(\gamma\))}  \\
    	\texttt{error},    & \text{otherwise}
	\end{cases}
\end{equation}

These guarantees expose a smaller attack surface for the implementing network, compared to the conventional general purpose protocol. General purpose protocols (implementing turing complete computation environments) aim to enable users to create an infinite amount of computations. The difference with this proposal is the amount of computations is bounded above by the protocol developer. For any protocol that accepts all computations, the attack surface can be as large as the acceptable computation space. Bounding the computation space above guarantees that the protocol will only accept a subset of the computation space, of which is decided upon by the implementing developer.

This properties of the system ensure developers can safely deploy public-facing, permissionless, purpose-built cryptoeconomies. If a developer wants to build a network to support decentralized elections, the developer should not have to concern themselves with whether or not a nefarious actor will submit a rogue transaction, and attempt to exploit a bug written in a previous transaction. Furthermore, the developer does not need to worry about their network computing upon transactions that perform other behaviors outside the scope of the intended election-based behaviors decided upon. In essence, each chain holds its own transaction types.

Theoretically, this approach can ensure an attack similar to the famous "DAO Hack" cannot occur within the network -- unless explicitly, and mistakenly allowed by the implementing developers. Developers can still make mistakes during implementation, but adequate developer tooling can mitigate those issues.

Turing complete language-driven, general purpose blockchains, are susceptible to an infinite amount of attack surfaces, because turing complete languages with no restrictions can be used to develop an infinite amount of malicious sets of code. 

This is largely due to the size of the Language \(L\) that can be represented using a turing machine. With any finite, non-empty, alphabet such as \(\texttt{A} = \{a, b\}\) there are an infinite number of finite-length words that can potentially be expressed: \("a", "aab", "bbabba", "aabbaabbaa"\), etc. 

Thus, formal languages are usually infinite, and describing an infinite formal language is not as simple as writing \(L = \{"cat", "dog", "catdog","dogcat"\}\). 

The attack surface of a system is the sum of the various points where an unauthorized user can attempt to attack an environment. For a language, the attack surface is as infinite as \(L\) itself. We propose simply bounding the attack surface above.

\begin{theorem}
	Let  \( \mathcal{S} \) be the set of attack surfaces for a given application \( \mathcal{A} \). If the application is designed to accept an infinite computations \( \mathcal{C} \), any set of restrictions \( \mathcal{R} \) placed upon the application, restricting a positive number of computations, \( \mathcal{R}(\mathcal{A}) \), thereby decreasing the amount of acceptable computations, decreases cardinality of \( \mathcal{S} \)
\end{theorem}

in other words,

\begin{equation}
	\mathcal{R}(\mathcal{A}) \rightarrow \vert\mathcal{S}(\mathcal{A})\vert  = \vert\mathcal{S}(\mathcal{A})\vert - \epsilon	
\end{equation}

where \(\epsilon\) is the number of computations no longer accepted after restriction. It is not the language that provides a smaller attack surface, it is the result of designing a purpose-built blockchain that accepts a subset of all turing-acceptable computations that yields a smaller attack surface.

\begin{equation}
	\begin{cases} 
      	\texttt{if } \epsilon > 0 &  \vert\mathcal{A}\vert > \vert\mathcal{A}\vert - \epsilon \\
      	\epsilon = 0   &  \vert\mathcal{A}\vert = \vert\mathcal{A}\vert \\
   	\end{cases}
\end{equation}

We also assume restricting inputs to \(\mathcal{A}\) do not result in more attack surfaces. Platforms such as Ethereum maintain documentation on how to further "secure" a blockchain (restricting certain accounts, etc), and it is growing over time. However, we can render smaller attack surfaces by design fundamentals, and not additional security measures entrusted upon the developer.

\chapter{Graph-based Analytic Engine}

\section{Query Graph}

\begin{figure*}[ht]
\centering
	\includegraphics[width=\textwidth]{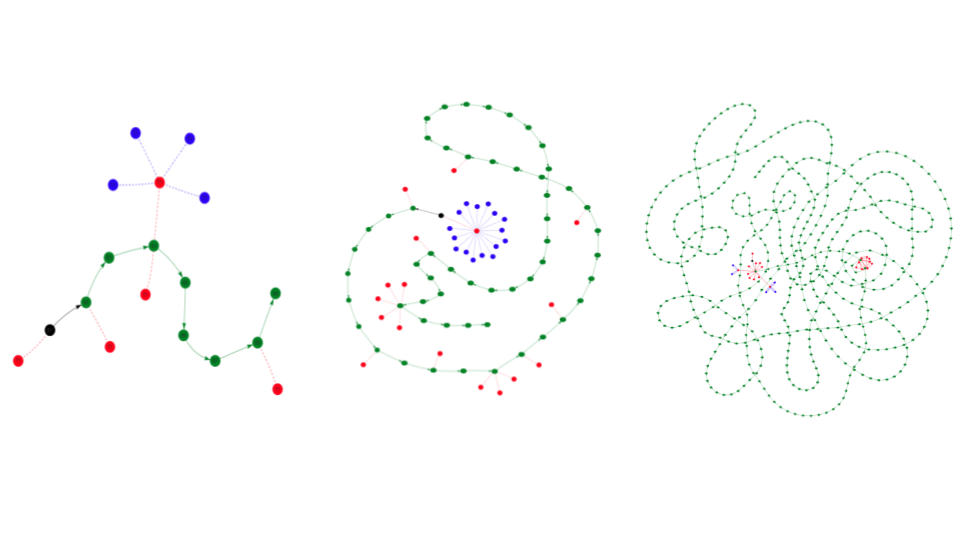}
	\caption{An illustration of how we represent an arbitrary \( \boldsymbol{\mathcal{B}} \) as network graphs, each chain varying in block height (left-to-right). This data structure is used to query the chain. Black is the genesis block, \(B_0\), green is any subsequent block, \(B_{0+}\), red is \(\gamma_f\), and blue is \(\gamma_m\). }
	\label{3_networks}
\end{figure*}

A graph database is any storage system that provides \textbf{index-free adjacency}. We use an ordered key-value store for root instance relationship storage, and propose its use for any relational data to be extracted from a chain. By using hexastores as the fundamental data store for each , we can further optimize relationship queries.

Current protocols make use of external, but node-local storage to retrieve context, and additional information regarding the state of the world. This is a common practice across protocols, so we purpose local storage for strategic technological advantages. In the case of relationship-based querying upon a given \(\boldsymbol{\mathcal{B}}\), a Graph model provides specific benefits, but is not necessary for operation, and should be ultimately left up the designer of the protocol.

Since we know the volume of transactions can be high in a blockchain network, in the service of efficient search, and querying, we store all processed root instances, discussed in section \ref{accbase}, in a graph database. One thing that remains, is the need for querying the protocol to determine the state of any given root instance. 

When importing data into a graph database, the relationships are treated with as much value as the database records themselves. This allows the engine to navigate connections between nodes in linear time, with respect to the tree depth required for traversal. That compares favorably to the exponential slowdown of many "Join SQL" queries in a traditional relational database.

\subsubsection{Graph Model}
We make use of the Subject \(S\), Predicate \(P\), Object triple store. This can be generalized to include n-tuple stores. \(S,P, and O\) are configurable in more advanced settings. Initially, we make use of the following predicates for proof-of-concept:

\begin{enumerate}
	\item \texttt{\textbf{Targeted}}
	\begin{enumerate}
		\item Involves a transaction's "to" field. If Alice sends value to Bob, \(A \rightarrow Targeted \rightarrow B\) is the graph entry
	\end{enumerate}
	\item \texttt{\textbf{Created}}
	\begin{enumerate}
		\item Refers to the causer of the transaction, or the from field, \(A \rightarrow Created \rightarrow T\)
	\end{enumerate}
	\item \texttt{\textbf{StoredIn}}
	\begin{enumerate}
		\item  Uses the transaction hash as the subject, and the block number as the object, \(\texttt{H}(T) \rightarrow StoredIn \rightarrow B_{number}\)
	\end{enumerate}
	\item \texttt{\textbf{Mined}}
	\begin{enumerate}
		\item \(Mined(T)\) , reflects what node mined the block containing the transaction T, \(Node_1 \rightarrow Mined \rightarrow B_{number}\)
	\end{enumerate}
\end{enumerate}

The content in the triple stores is configurable for the implementing developer. To represent A to B, through C, we write, \(A \rightarrow C \rightarrow B\). For look-up efficiency, we can store different possible permutations of this relationship, and result in 6 unique triple stores accessible by key: 

We store 6 keys for each triple:
\begin{enumerate}
	\item \texttt{spo}: \(A \rightarrow C \rightarrow B\)
	\item \texttt{sop}: \(A \rightarrow B \rightarrow C\)
	\item \texttt{ops}: \(B \rightarrow C \rightarrow A\)
	\item \texttt{osp}: \(B \rightarrow A \rightarrow C\)
	\item \texttt{pso}: \(C \rightarrow A \rightarrow B\)
	\item \texttt{pos}: \(C \rightarrow B \rightarrow A\)
\end{enumerate}

For any Search function, on any graph, \(\mathcal{G}\) , we simply use \(\texttt{search}(\mathcal{G})\). For example to search for the existance of \(T\) in \(\mathcal{G}\), we write 

\begin{equation}
	\texttt{search}(\mathcal{G}, T) \equiv
	\begin{cases}
		\texttt{true} & \texttt{if } T \in \mathcal{G} \\
		\texttt{false} & \texttt{otherwise}
	\end{cases}
\end{equation}

\section{Analytic Engine} 
We include an analytical layer of capabilities in the architecture for the purpose of native, quantitative decision-making. 

For example, if we want to determine the average block time for a given \( \boldsymbol{\mathcal{B}} \), we can create a vector, \(\mathcal{J}\), consisting of the timestamp, \(t\), differences between each block, \(b_{h_{t}}\)

\begin{equation}
	\frac{\Sigma_{1}^{j} }{\vert \mathcal{J} \vert} , \forall j \in \mathcal{J} , j \equiv b_{j+1_{t}} - b_{j_{t}}
\end{equation}

Through the Analytic Engine, and programming language, we expose many estimators for the purpose of decision making for developers.

\subsection{Inter-Economy Analytics}
With a common system powering many separate Blockchain economies, the opportunity arises to provide analytic solutions that span across several unique blockchain-based ecosystems. For example, to find a given transaction, \(T\), we can exploit fundamental search functions such as,

\begin{equation}
	\texttt{contains}(T, ((\boldsymbol{\mathcal{B}}_1 \cup \boldsymbol{\mathcal{B}}_2) \setminus \boldsymbol{\mathcal{B}}_3))
\end{equation}

At the time of writing, the application of statistical analysis to cryptoeconomic systems is a growing field of research.

\chapter{Native Abstraction Language}

\section{Formal Language}
To create a developer experience that resembles actual programming, we expose an expressive, turing-incomplete formal language purposed with allowing developers to represent their \( \boldsymbol{\mathcal{B}} \) specification as a programming language. This language also allows developers to interact with the chain, once active. This creates one common, fundamental language for designing a blockchain protocol, and interacting with it. 

\subsection{Defining Data Types}
A main goal of this work is to take developer experience into account in the highest regard.  

\subsection{Data Type}
The Integer Data Structure, for example is represented a specific way, and you can call operations on it. This can vary based on the programming language in use, but the notion of taking a concept such as an Integer, and allowing higher level functions to be invoked upon it inspires the following data types. 

\subsection{Blockchain Data Type}
To configure a blockchain, the language expects usage of the \(\texttt{Blockchain}\), and \(\texttt{Consensus}\) symbols. These provide the functionality needed to construct a blockchain design. The \(\texttt{Root}\) keywork is required for tasks such as adding a new root.

The simplest example, while still showing programmatic capabilities, could be code snippet \ref{code:simple},  

\begin{listing}[ht]
\begin{minipage}{\linewidth}
\begin{lstlisting}

Blockchain B1(Consensus) {

	this.consensus = Consensus.POW;

	func Create(Config i, Status s){ 
		log("created..."); 
		return True; 
	}; 

	func testFunc(Block b){
		Nonce answer = (b.nonce);
		log(answer);
	}

	func OnNewBlock(Block b, Hash h){
		log("Block ID: "+b.id);
		log("Block Hash: "+h);
		Int number_result = testFunc(b);
		log(number_result);
	}

}
\end{lstlisting}
\end{minipage}
\caption{Simple chain creation}
\label{code:simple}
\end{listing}

Since a chain is a data type, we can invoke upon it. To submit a new root instance, we can call send, and pass our root instance as a parameter.

\begin{equation}
\texttt{B1.send(\ldots)}
\end{equation}

To process a condition on whether a given root instance exists in a chain, we can reference all on chain root instances, \(\texttt{RI}\), and check for the existence of one.

\begin{equation}
\texttt{if(B1.RI.contains(\ldots))\{\ldots\}}
\end{equation}

For the task of building a chain dedicated to elections, we must configure the chain to accept new ballots, and cast new votes. This is demonstrated in code snippet \ref{code:election_chain},

\begin{listing}[ht]
\begin{minipage}{\linewidth}
\begin{lstlisting}
import ballot;
import verdict;

Blockchain election_chain(Consensus, Roots) { 

    this.consensus = Consensus.POW;

    Roots.add(ballot);
    Roots.add(verdict);
	
    func Create(){ 
        log('created...'); 
    }; 

    func OnNewBlock(){
        log("new block...");
    }

}
\end{lstlisting}
\end{minipage}
\caption{Election chain creation}
\label{code:election_chain}
\end{listing}

At the time of writing there does not exist a high-level programming language specifically for the designing of a blockchain protocol. With such an expressive language, we can lessen the blockchain development learning curve. Assuming such a language does not instead make it more difficult for developers to use. The aim of this language is to be the only mechanism developers use to create, and interact with a given blockchain.

\subsection{Root Type}
Root represents objects to be tracked on the chain,  by way of a tree containing the objects. Each Root can have multiple Aspects used by it.

\begin{listing}[ht]
\begin{minipage}{\linewidth}
\begin{lstlisting}
import votes;

Root root_name(){
	AddAspect(votes)
	...
}
\end{lstlisting}
\end{minipage}
\caption{Root example for casting a vote}
\label{code:2}
\end{listing}

\subsection{Aspect Type}
Aspects define variable types within Roots. The aspect's root, and other roots can access the aspect if allowed.

\begin{listing}[ht]
\begin{minipage}{\linewidth}
\begin{lstlisting}
Aspect votes{
	description = "..."
	default_value = 0
}

\end{lstlisting}
\end{minipage}
\caption{Root aspects for casting a vote}
\label{code:4}
\end{listing}

\subsection{Mechanism Type}
We use the Mechanism type to design algorithmic processes for the network to follow. We can not only construct consensus mechanisms with this, but additional network-related rules and processes. For example, we can represent proof of work by using the mechanism type.

\begin{listing}[ht]
\begin{minipage}{\linewidth}
\begin{lstlisting}
Mechanism proof_of_work{
	SocialWelfare(){...}
	SocialChoice(){...}
	Valuation(){...}
	...	
}
\end{lstlisting}
\end{minipage}
\caption{The structure of a mechanism, including several decision functions}
\label{code:5}
\end{listing}

By applying the mechanism to the chain configuration, we put it to use. However, in addition to the fundamental consensus property of the chain, we can also define, a additional rule. Each mechanism has native methods, each triggered at different times. For example, the \(\texttt{Execute}\) method is invoked every time the mechanism is invoked.

\begin{listing}[ht]
\begin{minipage}{\linewidth}
\begin{lstlisting}
Mechanism say_hello{
	Execute(){
		log("hello");
		...
	}
}
\end{lstlisting}
\end{minipage}
\caption{A native mechanism example with a triggered function}
\label{code:6}
\end{listing}

The \(\texttt{Execute}\) function is native, and is invoked on when the mechanism is executed; this too can be configured.

\subsubsection{Mechanism Function Types}
Each mechanism can implement several functions to accomplish various goals. 

Upon a peer receiving a message from another, supposed a developer wanted to invoke a simple condition that compared a scalar value from each peer to decide what execution path the protocol should take.

\begin{listing}[ht]
\begin{minipage}{\linewidth}
\begin{lstlisting}
Mechanism ScalarCompare{
	OnPeerMessage(peer){
		if (peer.message > 1){
			Broadcast("hello")
		}
	}
}
\end{lstlisting}
\end{minipage}
\caption{Mechanism for ScalarCompare}
\label{code:7}
\end{listing}

With these paradigms, developers can build several custom mechanisms into their deployed protocol specification.

\subsection{Puzzle Data Type}
Using a native puzzle data type, we can abstract the creation of computations of which participants can attempt to solve to further the creation of an economy. This is arbitrary, and serves as an example of the types of abstractions we can expose during the developer experience of constructing, and designing blockchains.

\subsection{Native Functions}

By design, developers declare functions to be ran during several scenarios, all of which are included in the genesis block for a chain. This provides a single source of instructions to be computed. For example,

\begin{listing}[ht]
\begin{minipage}{\linewidth}
\begin{lstlisting}
func OnNewBlock(...){
	...
}
\end{lstlisting}
\end{minipage}
\caption{Chain function for \(\texttt{OnNewBlock}\)}
\label{code:8}
\end{listing}

is invoked upon every new block created in the network. Each node runs this function if they create a new block. Other native functions include, but are not limited to,

\begin{enumerate}
	\item \texttt{OnCreate}
	\item \texttt{OnNewPeer}
	\item \texttt{OnBlockReceived}
\end{enumerate}

\section{Chain Functions}
Each configuration file from which a chain is constructed can contain sets of instructions to be computed during several scenarios during the life of the chain. \texttt{OnNewBlock} is but one example of this, and can be arbitrary. Chain functions differ from smart contracts because chain functions are declared during chain creation time, and serve as "hard-coded" computations of which occur at different times. Chain functions help dictate the behavior of a chain, opposed to being declared by a smart contract. Smart contracts serve as the providers of "input" into chain functions already declared in a chain's genesis. For security purposes, chain functions cannot be rewritten once a chain is deployed into the "wild".

\chapter{On Protocol Properties}
 
\section{Protocol Properties}

\subsection{Node Transparency}
Decentralization is a core aspect of a Blockchain ecosystem, however the purpose of such an aspect is important to consider. Decentralization allows nodes to arrive at agreement, with no central point of failure. Distributing the nodes across the globe is not the only way to achieve decentralization. For example, you can have a fully decentralized blockchain economy of which only exists within one physical building. Ownership comes into play when determining the decentralization of any network. This property assumes the data, and the logic will exist on each node. Regardless on the implementation, transprency of the data on a blockchain should never be sacrificed. Moreover, it may not be necessary to exclusively couple logic, and data onto every node.

\subsection{Logic Separation} \label{logicsep}
Blockchain ecosystem, since Bitcoin, have implemented systems within which network nodes can either store all of the data, or a part of it. In Bitcoin, these partial nodes are use the Simplified Payment Verification (SPV) mechanism for validation of transactions. The proposal includes nodes that only carry the data of the blockchain. Using Bitcoin as an example, the "data" in this description can represent the "transactions" of their system. 

The proposed system can separate Logic and Data, between nodes, keeping all capabilities on each node, or enable a simplified mode similar to SPV. Separating Logic and Data allows for not only a separation of concerns but also for more transparency in systems that may be more centralized than a pure decentralized implementation.

We draw an analog to being "judged", or "the judger". Using this design, some nodes (preferably nodes controlled by sole entities) can be purposed with performing logic on data, while public nodes, can maintain the data of a system. This relinquishes data-nodes from being concerned with heavy computation, while still enabling public transparency into the state of a Blockchain ledger. Cryptoeconomically, we can incentivize data-nodes, and logic-nodes in different ways, of which we can yield to the developers, if inquired to do so. We can also have a hybrid approach where logic-nodes can store the data as well, peered with data-nodes to keep them "honest". Once again, these decisions are far too complicated to generalize, and should be decided by the implementing developer.

A concern arises regarding any logic-based node simply executing logic, and can introduce a vulnerability. We can mitigate this, by design, by enabling logic-based nodes to simply hold representations of the "source-of-truth", the blockchain ledger, but only executing logic on the data passed to it, but using it's own data to validate they are equal.

\subsection{Programmatic Usage Conventions}
This article also proposes a simple developer interaction such as,

\begin{equation}
\texttt{Blockchain b = new Blockchain(\ldots)}
\end{equation}

if we focus on an object oriented approach. This allows any object oriented application to integrate with a fully-functional blockchain. The proposed implementation would only need network dependencies, as the system can be deployed in an enterprise manner. Considering a functional approach can be,

\begin{equation}
\texttt{var b = Blockchain(\ldots)}
\end{equation}

Both of these developer interactions aim for the least developer-friction, with no sacrafice in decentralization, scalability, and security.

\subsubsection{REST API}
the system can also expose a REST-like Application Programming Interface (API), for a more developer friendly experience. It can be arbitrary regarding what functionality to expose through an API, but should be of use to the developer invoking the protocol in use.

\subsection{Decentralized Nodes}
When Bitcoin first began to propagate, anyone could mine on their computer, due to the ease of mining at that time. Users of whom wanted to maintain a wallet could do so on their computer as well. Over time, many developers, and users of Bitcoin have begun to maintain their wallets on a cloud server. At the time of writing, a large majority of "Segregated Witness" Bitcoin nodes exist in Amazon Web Service instances (servers). There does currently exist an appetite, and market for "cloud mining" but has shown to be less profitable when compared to dedicated mining equipment; this may be due to the fees charged for cloud mining services.

It has yet to be definitively shown whether cloud mining will be beneficial for users, as it also depends on the protocol being mined, but users maintaining their version of a blockchain's history has already proven to be useful, and has been a natural evolution of cryptocurrencies. This is because maintaining a wallet does not require large amounts of computing power, or storage that isn't readily available on consumer grade hardware. Rightfully so, this also depends on the protocol being represented by the wallet. Because of this, we experiment with a semi-centralized architecture, giving developers the option to choose hosting types, and rapidly experiment.

\subsubsection{Hosted Blockchain Economies}
As of the time of this writing, cloud computing providers now offer services to fully host specific protocol platforms, mainly Hyperledger, and Ethereum. These services simply attempt to abstract away most of the implementation details for these platforms. This does not allow the developer to customize the properties of the network beyond the arbitrary variables implemented by the protocol of choice. This is a step in the right direction, but it is ultimately on the shoulders of the protocol developers to enable protocol customization, and further simplification.

As covered in section \ref{logicsep}, we can provide options to a developer regarding where they would like their nodes to exist, without sacrificing the decentralization property required for any cryptoeconomic environment. Similar to Bitcoin's "Full", and SPV node types, we can tier the requirements for nodes to maintain the chain data locally. However, we can do the same for computation privileges. Simplified nodes can be the nodes holding the data to be computed on, while full nodes can hold reference to it, but not it. 

We can enable the option of allowing full nodes to keep a record of the data, with which to verify incoming data from a "data node", but not to compute upon it. Simplified nodes can also choose to keep a record of the data, from which full nodes do not compute, or verify; this allows for simply keeping a synced record of the blockchain ledger used. 

Developers should also be able to choose if they are indifferent to whether they need to maintain any of the blockchain data/operations themselves. Either way, the blockchain in question should have the right tooling to verify its cryptoeconomic properties are indeed holding true over time. There are a subset of developers of which do not care to maintain, verify, or manage the implementation of a Blockchain, and another subset that want to make design decisions on the protocol itself, but do not care to maintain it.

\chapter{Protocol Adaptation}

\section{Updating Code}
With systems built upon arbitrary blockchain designs, they face the difficult challenge of updating the code existing on the nodes in the wild. This is famously difficult, and results in the platform developers requiring a iterative-waterfall approach to their development releases. Developers must attempt to predict the behavior of network participants, in attempts to "get it right". This article proposes a system to simply store the version of a given blockchain ledger, but also update the blockchain rules, allowing for transitioning between code versions, without inconveniently interrupting, or disturbing network operations and properties.

\subsection{Rollovers}
To rollover an existing chain onto a new version of the same chain, we store a copy of the old chain, locally or remote, and hash its contents, \(\texttt{H}(\boldsymbol{\mathcal{B}})\). We then include the hash of the previous version's Blockchain into the genesis block of the new version. To hash a small chain, you can directly hash the chain as a single data structure. However, for larger chains, it depends on how the chain is stored. For a chain stored by several files on a system, we can hash each file, and construct a tree, finally hashing it's root. This serves as the same "proof" of existence at the time of hashing.

\section{Transactions per second}
To alter transaction per second, we can, among other things, decrease block mine/forge time, or decrease block size. What a network can achieve depends on several aspects. First generation blockchain systems are constrained at the network level because a large part of the network participates in transactions; this is different from the capacity of a single node. If the network is viewed as a graph, \(\mathcal{G}\), the diameter of the network influences how information propagates within it. For a single transaction, the amount of time it takes to span all nodes from inception to finality also influences this.

\chapter{Enterprise Deployment}

\section{Sole Enterprise Blockchain Usage}
For a large corporation, that may not be interested in forming, or joining a consortium, implementing a meaningful Blockchain ecosystem may not make sense by today's standard. However, instead of deciding that using a Blockchain doesn't apply, we can enable such a company with a cloud hosted blockchain economy for experimentation. This enables an on-going, ever-present blockchain network, upon which to store data, and from which to fetch data. 

\section{State}
"State" in bitcoin represents the state of the amount of money eberyone has, Ethereum uses state as how much money everybody has, what is the code for smart contracts, and what is the state of all of the smart contracts.

\section{Development Operations}
By considering developer experience, we can allow developers to choose how much control they want over an individual node, or the entire blockchain network. For developers who choose controlling their node, there exists plenty of platforms in existence today. For developers who want to deploy a blockchain protocol with minimal concerns regarding its configuration, we can abstract away the need for protocol design, and enable developers to invoke/interact with an arbitrary blockchain, of their creation, by simple programmatic syntax; this syntax can differ depending on the programming language used, but fundamentally, it can be of a functional nature, or fully object oriented.

\section{Development Interface}
For developers of whom want control over every aspect of the protocol's design, we can extend an interface, be it user interface based, or programmatic, to configure several aspects of the protocol's design properties. Creating a developer interface is different from open sourcing a project. Open sourced software, to extend its usefulness, inherently requires the extending developer to edit the code of the platform. Implicitly, this means the more complicated, robust a platform's code base becomes, the higher the effort becomes on the part of the extending developer. For these reasons, we choose a purposed developer interface.

\end{document}